\documentclass[aps,prx,preprint,superscriptaddress]{revtex4-1}

\usepackage{graphicx}

\begin{document}

\renewcommand{\figurename}{FIG.}

\title{Anomalous modulation of Josephson radiation in nanowire-based Josephson junctions}

\author{H. Kamata}
\thanks{These authors contributed equally to this work}
\email{hiroshi.kamata@riken.jp}
\affiliation{Center for Emergent Matter Science, RIKEN, 2-1 Hirosawa, Wako, Saitama 351-0198, Japan}
\affiliation{Department of Applied Physics, University of Tokyo, 7-3-1 Hongo, Bunkyo, Tokyo 113-8656, Japan}
\author{R. S. Deacon}
\thanks{These authors contributed equally to this work}
\affiliation{Center for Emergent Matter Science, RIKEN, 2-1 Hirosawa, Wako, Saitama 351-0198, Japan}
\affiliation{Advanced Device Laboratory, RIKEN, 2-1 Hirosawa, Wako, Saitama 351-0198, Japan}
\author{S. Matsuo}
\affiliation{Department of Applied Physics, University of Tokyo, 7-3-1 Hongo, Bunkyo, Tokyo 113-8656, Japan}
\author{K. Li}
\affiliation{Beijing Key Laboratory of Quantum Devices, Key Laboratory for the Physics and Chemistry of Nanodevices, and Department of Electronics, Peking University, Beijing 100871, China}
\author{S. Jeppesen}
\affiliation{Division of Solid State Physics, Lund University, Box 118, SE-221 00 Lund, Sweden}
\author{L. Samuelson}
\affiliation{Division of Solid State Physics, Lund University, Box 118, SE-221 00 Lund, Sweden}
\author{H. Q. Xu}
\affiliation{Beijing Key Laboratory of Quantum Devices, Key Laboratory for the Physics and Chemistry of Nanodevices, and Department of Electronics, Peking University, Beijing 100871, China}
\affiliation{Division of Solid State Physics, Lund University, Box 118, SE-221 00 Lund, Sweden}
\author{K. Ishibashi}
\affiliation{Center for Emergent Matter Science, RIKEN, 2-1 Hirosawa, Wako, Saitama 351-0198, Japan}
\affiliation{Advanced Device Laboratory, RIKEN, 2-1 Hirosawa, Wako, Saitama 351-0198, Japan}
\author{S. Tarucha}
\email{tarucha@ap.t.u-tokyo.ac.jp}
\affiliation{Center for Emergent Matter Science, RIKEN, 2-1 Hirosawa, Wako, Saitama 351-0198, Japan}
\affiliation{Department of Applied Physics, University of Tokyo, 7-3-1 Hongo, Bunkyo, Tokyo 113-8656, Japan}

\date{\today}

\begin{abstract}
We investigate the Josephson radiation of nanowire (NW)-based Josephson junctions in a parallel magnetic field.
The Josephson junction made of an InAs NW with superconducting Al leads shows the emission spectra which follow the Josephson frequency $f_{J}$ over the range 4-8~GHz at zero magnetic field.
We observe an apparent deviation of the emission spectra from the Josephson frequency which is accompanied by a strong enhancement of the switching current above a magnetic field of $\sim 300$~mT.
The observed modulations can be understood to reflect trivial changes in the superconducting circuit surrounding the device which is strongly affected by the applied magnetic field.
Our findings will provide a way to accurately investigate topological properties in NW-based systems.
\end{abstract}

\maketitle

Topological properties in condensed matter systems have attracted much attention for developing fault-tolerant topological quantum computation \cite{nayak:2008}.
In particular, InAs or InSb semiconductor nanowires (NWs), which exhibit a strong spin-orbit interaction and a large Land\'e g-factor, coupled to a s-wave superconductor are good candidates for topological superconductors that can host non-Abelian quasi-particles, the so-called Majorana fermions (MFs) forming the basis of topological quantum computation \cite{kitaev:2001, lutchyn:2010, oreg:2010}.
MFs have been primarily investigated in tunneling spectroscopy measurements of zero bias conductance peaks on normal conductor/NW/superconductor hybrid junctions \cite{mourik:2012, das:2012, deng:2016, zhang:2017, nichele:2017, suominen:2017}.
Further experimental evidence of the MFs is found in the observation of a 4$\pi$-periodic Josephson effect in a topological Josephson junction \cite{beenakker:2013}.
Fusion of two MFs produces an ordinary fermion and modifies the periodicity of the Josephson relation from 2$\pi$ (Cooper pairs) to 4$\pi$ (MF pairs) with protected crossings at a superconducting phase difference $\phi = \pi$ \cite{kwon:2004}.
In NW-based Josephson junctions, MF-related signatures have been observed in measurements of dc Josephson supercurrent \cite{tiira:2017}, Shapiro steps \cite{rokhinson:2012}, and Josephson radiation \cite{laroche:2017}.
Although these results are intrinsically related to each other, no systematic investigation has been reported.

In general, a finite voltage bias $V$ across a superconducting weak link generates the Josephson radiation with the universal Josephson frequency
\begin{equation}
f_{J} = \frac{2e}{h} V,
\end{equation}
linking the tunneling of Cooper pairs with a net charge of $2e$ to photons of energy $h f_{J}$, where $e$ is the single electron elementary charge, $h$ is the Planck constant.
In contrast, radio-frequency (rf) excitation of frequency $f$ induces quantized voltage steps (Shapiro steps) with a height $\Delta V = h f / 2e$ in the dc current-voltage characteristics.
In topological Josephson junctions made of HgTe-based topological insulators \cite{konig:2007, brune:2011} with superconducting leads, a tunneling of quasiparticles with charge $e$ attributed to the presence of topological gapless Andreev bound states gives rise to radiation at the half the Josephson frequency $f_{J}/2$ \cite{deacon:2017} and the doubling of the Shapiro step size $h f / e$ \cite{wiedenmann:2016, bocquillon:2016}, being consistent with each other.
In NW-based Josephson junctions, on the other hand, a topological phase transition requires an external magnetic field, which generally modulates transport properties of NWs as well as superconducting thin films.
Therefore, to characterize topological properties of junctions, transport properties of not only the junction but also its surrounding circuit should be taken into account.

Here, we report direct rf measurements of emission spectra on NW-based Josephson junctions in a parallel magnetic field.
We find that the emission spectra deviate from the fundamental radiation accompanying strong enhancement of the switching current even in the trivial regime.
The observed deviation can be understood to reflect a modulation of the voltage bias across the junction in the finite magnetic field.
To confirm the interpretation we evaluate the actual voltage bias across the junction by measuring the Shapiro response under rf irradiation.
We note that in contrast to microwave spectroscopy measurements utilizing an on-chip detector \cite{woerkom:2017}, which is affected by the external magnetic field, our direct high frequency measurement \cite{deacon:2017} will be more appropriate to investigate ac Josephson effects in the high magnetic field regime or nontrivial topological regime, albeit with a significantly reduced detection bandwidth.

The Josephson junction with an InAs NW weak link and Al contacts was fabricated on a gate dielectric of hexagonal boron nitride (hBN).
A hBN dielectric (thickness of $\sim 30$~nm) made by exfoliating commercial BN powder (Momentive PT110) with Scotch tape was mechanically transferred onto local gate electrodes (Ti/Au) patterned on a Si substrate by means of a dry transfer technique \cite{wang:2016, baba:2017}.
A wurtzite InAs NW grown by chemical beam epitaxy along the $\langle 111\rangle$ direction with an average diameter of 80~nm and an average length of 2~$\mu$m was transferred onto the gate substrate with the same dry transfer technique as the case of hBN.
The relative position of the NW with respect to pre-defined gate electrodes is determined with the help of atomic force microscope images.
Electrical contacts to the NW with width of $\sim 200$~nm and spacing of $\sim 60$~nm are defined by conventional electron beam lithography.
As the Al contact width is shorter than the superconducting coherence length $\xi_{\rm Al} \sim 1.6$~$\mu$m, its critical field for the in-plane direction is enhanced \cite{moshchalkov:1995, poza:1998}.
In addition, an on-chip transmission line (1250~$\mu$m in length and 10~$\mu$m in width) connected to one of the contacts is defined in the same fabrication step as the contacts.
Before the deposition of superconducting Ti/Al (1.5~nm/90~nm) contacts, the NW is chemically etched in a (NH$_{4}$)$_{2}$S$_{x}$ solution \cite{suyatin:2007} followed by a brief {\it in-situ} low-power Argon plasma cleaning to remove the native surface oxide layer \cite{gul:2017, gul:2018}.
To minimize stray inductance and enable a stable and accurate voltage bias across the junction, an on-chip resistance made of Cr/Au (8~nm/8~nm) shunts the junction \cite{chauvin:2006}.
Different NWs prepared in the same way show clear quantized conductance as a function of a gate voltage applied to the local gate electrode underneath the junction (see Supplementary Information).

Figure~\ref{Fig1}(a) shows an optical micrograph of the device and experimental setup.
The junction is connected to a coaxial line through the on-chip transmission line for the emission measurement.
The microwave and dc measurement lines are decoupled via a bias tee.
The dc current-voltage characteristics were measured by applying a bias current $I_{sd}$ to the junction through the bias tee, a copper semi-rigid coaxial line and circuit board of the sample holder. 
A voltage drop $V_{sd}$ between the bias tee and superconducting ground plane of the device was measured.
The resistance due to the normal sections of the measurement circuit was evaluated from the resistance observed in the supercurrent branch ($R_{c} \sim 14$~$\Omega$) at $B = 0$~mT and is subtracted from all data sets.
The Josephson radiation emitted from the junction was amplified by cryogenic and room-temperature amplifiers and then measured with a swept spectrum analyzer \cite{deacon:2017}.
The commercial rf components used in the readout line limit the frequency range of detection to 4-8~GHz.
In addition, another coaxial line is coupled into the measurement circuit through a directional coupler to all rf-stimulation of the junction to observe Shapiro steps.
In-plane magnetic field $B$ is applied parallel to the NW axis.
All measurements were carried out in a dilution refrigerator at $\sim 30$~mK.

Figure~\ref{Fig1}(b) shows typical examples of the dc current-voltage characteristic (black line) and numerically calculated differential resistance $dV/dI$ (blue line) observed at $B = 0$~mT.
The observed switching current $I_{c} \sim 150$~nA without hysteretic behavior indicates the formation of an ideal overdamped junction, in which capacitance and resistance are small.
The resistance of $\sim 40$~$\Omega$ outside the superconducting state corresponds to the shunt resistance which dominates the circuit in the normal branch as the junction resistance is typically $\geq 500$~$\Omega$ (see Supplementary Information).
We note that the observed supercurrent was only weaker sensitive to gate voltage $V_{\rm G}$ applied to all local gates due to electric field screening caused by the narrow contact spacing of $\sim 60$~nm [inset of Fig.~\ref{Fig1}(a)].
Therefore, all measurements which follow were carried out at a fixed voltage of 0~V on all local gates.

We first confirm the ac Josephson effect in this device by measuring the Josephson radiation at $B = 0$~mT.
Figure~\ref{Fig1}(c) shows an emission spectrum observed at $B = 0$~mT as a function of bias current $I_{sd}$ with fixed detection frequency of $f_{d} = 7.01$~GHz.
A clear peak appears when the Josephson radiation matches the detection frequency $f_{d} = f_{J} = 2eV/h$.
Figure~\ref{Fig1}(d) shows emission spectra observed at $B = 0$~mT as a function of voltage bias $V_{sd}$ and detection frequency $f_{d}$ (the data is normalized to the maximum emission amplitude for each $f_{d}$).
The observed emission lines follow the linear relation of the fundamental Josephson frequency $f_{J} = 2eV/h$ over the range 4-8~GHz, indicating the conventional 2$\pi$-periodic Josephson effect.
In principle, the linewidths of emission lines reflect the lifetime of the Andreev bound states.
The observed line at $f_{J}$ exhibits a typical width of $\delta V \sim 5$~$\mu$V, yielding a coherence time of $t = h/2e \delta V \sim 0.5$~ns.
This value is generally consistent with the lowest drive frequency for which Shapiro steps were well resolved in this device ($\sim 2$~GHz).

We now investigate the dependence of the supercurrent [Fig.~\ref{Fig2}(a)] and the Josephson radiation [Fig.~\ref{Fig2}(b)] on the in-plane magnetic field $B$ applied parallel to the NW axis.
As shown in Fig.~\ref{Fig2}(a), as $B$ is increased, while the switching current $I_{c}$ remains almost constant up to $B \sim 250$~mT, $I_{c}$ is strongly enhanced at $B \sim 300$~mT, and monotonically decreases at higher $B$.
On the other hand, Fig.~\ref{Fig2}(b) shows Josephson radiation observed at fixed detection frequency of $f_{d} = 5.0$~GHz as a function of magnetic field $B$ and voltage bias $V_{sd}$ (the data is normalized to the maximum emission amplitude for each $B$).
As $B$ is increased, the observed emission spectra deviate from the Josephson voltage $V_{J} = (h/2e) f_{J} \sim 10.35$~$\mu$V at $B \sim 250$~mT and $\sim 300$~mT, coinciding with the $B$ dependence of the dc supercurrent.
We note that the amplitude of the emission signal is, in principle, proportional to the amplitude of the switching current $I_{c}$ observed in the dc measurement.
Although the emission signals are normalized for clarity, the observed increase in the linewidth of the emission spectra and its background noise at $B > 300$~mT indicate the strong suppression of the emission power, which is not consistent with the enhancement of $I_{c}$ [Fig.~\ref{Fig2}(a)].

To look into the details of the abrupt changes, in Fig.~\ref{Fig2}(c) we plot the differential resistance $dV/dI$ as a function of bias current $I_{sd}$ observed at $B = 0$~mT, 270~mT and 340~mT.
The resistance in the entire range of $I_{sd}$ increases at $B \geq 270$~mT.
In addition, Fig.~\ref{Fig2}(d) shows $B$ dependence of $dV/dI$ in the superconducting branch ($I_{sd} = 0$~nA) and normal branch ($I_{sd} = 600$~nA).
The normal branch resistance, which is dominated by the shunt resistance placed in close proximity to the junction, also shows an abrupt increase by $\sim 15$~$\Omega$ at $\sim 250$~mT.
Therefore, the observed deviation of the emission spectra [Fig.~\ref{Fig2}(b)] is considered to reflect the a change of the resistance outside the shunt circuit rather than a dissipative component coming from the junction inside the shunt circuit.

To identify the origin of the observed deviation in the emission spectra, we now turn to measurements of Shapiro steps under rf irradiation.
Shapiro response provides a convenient probe for the periodicity of the Josephson supercurrent as well as the actual voltage drop across the junction.
We observe that the positions of Shapiro peaks clearly require correction due to an extra series resistance \cite{shapiro:1963}.
Figure~\ref{Fig3}(a) shows two-dimensional color plots of the bin counts of each Shapiro step as a function of voltage bias $V_{sd}$ (in normalized units $hf/2e$) and rf excitation power $P_{\rm rf}$ observed at fixed frequency of $f = 4.0$~GHz at $B = 0$~mT.
This plot clearly indicates the conventional 2$\pi$-periodic Josephson effect in this junction, exhibiting a quantized voltage $V_{n} = nhf/2e$.
On the other hand, Fig.~\ref{Fig3}(b) shows $B$ dependence of the bin counts at fixed frequency of $f = 4.0$~GHz and power of $P_{\rm rf} = 1.0$~dBm.
Each Shapiro peak deviates from the quantized value of $nhf/2e$ at $B \geq 250$~mT, indicating the emergence of an additional voltage drop in other parts of the circuit.

We correct the deviated Shapiro steps at $B \geq 250$~mT by considering an effective resistance $R_{eff}$ in the circuit.
Figure~\ref{Fig4}(a) shows a schematic circuit diagram in this measurement.
We attribute an anomalous resistance $R_{anom}$ only to the transmission line (1250~$\mu$m in length and 10~$\mu$m in width) because branches connecting the transmission line to the shunt resistance (35~$\mu$m in length and 5.5~$\mu$m in width) and the Josephson junction (35~$\mu$m in length and $\sim 5$~$\mu$m in average width) are comparatively small in volume compared with the transmission line and the resistance of the shunt itself [Fig.~\ref{Fig1}(a)].
The solid line in Fig.~\ref{Fig4}(b) shows $B$ dependence of the anomalous resistance $R_{anom}$ (estimated from $B$ dependence of the normal branch resistance in Fig.~\ref{Fig2}(d) subtracting that at $B = 0$~mT) plus the constant normal resistance in the measurement circuit of $R_{c} = 14$~$\Omega$ due to the bias tee, semi-rigid coaxial lines and circuit board of the sample holder.
Figure~\ref{Fig4}(c) shows a corrected Shapiro response observed at $B = 340$~mT with $R_{eff} = 29.8$~$\Omega$.
This plot exhibits a quantized voltage $V_{n} = nhf/2e$ and reveals the conventional, not 4$\pi$-periodic, Josephson effect.
We note that in both low power and high power oscillatory regimes we observe the same sequence of Shapiro steps in contrast to recent observations of doubled steps in the low power regime \cite{wiedenmann:2016, bocquillon:2016, deacon:2017}.
As shown in Fig.~\ref{Fig4}(b), the effective resistance $R_{eff}$ (circles) used for the correction is consistent with $R_{anom} + R_{c}$ (solid line) for all magnetic fields.
Therefore, we conclude that the Shapiro features observed for all magnetic fields reveal a conventional 2$\pi$-periodic Josephson effect.
As a result, $B$ dependence of a corrected Josephson radiation exhibits conventional emission spectra even at $B \geq 250$~mT as shown in Fig.~\ref{Fig4}(d).

We attribute the observed modulations [Figs.~\ref{Fig2}(a) and \ref{Fig2}(b)] to a non-topological change of the surrounding circuit in the applied magnetic field caused simply by sections of the Al leads undergoing a transition to the normal state, while the junction contacts remain superconducting.
As shown in Fig.~\ref{Fig1}(a), the junction has a remote four-terminal measurement setup where the four-terminal connections rely on the circuit made of Al thin films (thickness of 90~nm) being a superconducting state.
Applying an in-plane magnetic field, superconductivity in the circuit is suppressed at $B > 300$~mT \cite{reale:1974}, and then, an additional series resistance emerges.
In addition, small modulations observed at $B = 250 \sim 300$~mT may be caused by the same behavior of Al bonding wires which connect the device and the electronic circuit inside the refrigerator.
We note that when Au bonding wires were employed, such a small modulation in the Josephson supercurrent was not observed (see Supplementary Information).
On the other hand, superconductivity on the junction remains even at $B > 300$~mT owing to the small size of contacts as compared to the superconducting coherence length $\xi_{\rm Al} \sim 1.6$~$\mu$m \cite{moshchalkov:1995, poza:1998}.
The observed enhancement of $I_{c}$ [Fig.~\ref{Fig2}(a)] may reflect a filtering of high-frequency noise on the transmission line which turns to the normal state at $B > 300$~mT.
Although quantitative discussion requires further investigation, our results clearly indicate that non-topological change of the surrounding circuit modulates the Josephson radiation as well as the Josephson supercurrent above a finite magnetic field.

In summary, we have investigated the Josephson radiation of an InAs NW junction with Al superconducting leads in a parallel magnetic field.
We observe anomalous modulation of the Josephson radiation and strong enhancement of the switching current above the critical field of Al thin films.
The observed modulation of the Josephson radiation can be explained by a change of the circuit impedance.
In addition, the observed enhancement of the switching current can be understood by considering a filtering of high-frequency noise on the circuit.

In primarily measurements of MFs using tunneling spectroscopy, which are performed in a highly-resistive regime \cite{mourik:2012, das:2012, deng:2016, zhang:2017, nichele:2017, suominen:2017}, the trivial changes of the surrounding circuit can be neglected.
However, our results may modify the interpretation for recent measurements of NW-based Josephson junctions \cite{tiira:2017, laroche:2017}, where a voltage bias across the junction plays a crucial role.

\section*{Acknowledgments}
This work was partially supported by JSPS Grant-in-Aid Scientific Research (A) (No.~JP16H02204), JSPS Grant-in-Aid for Scientific Research (S) (No.~JP26220710), ImPACT Program of Council for Science, Technology and Innovation (Cabinet Office, Government of Japan), and Grant-in-Aid for Scientific Research on Innovative Areas ``Science of Hybrid Quantum Systems" (No.~JP15H05867) from MEXT.
H.K. acknowledges support from RIKEN Incentive Research Projects, and JSPS Early-Career Scientists (No.~JP18K13486).
S.M. acknowledges support from JSPS Grant-in-Aid for Young Scientific Research (A) (No.~JP15H05407), JSPS Grant-in-Aid Scientific Research (B) (No.~JP18H01813), Grant-in-Aid for Scientific Research on Innovative Area ``Topological Materials Science" (No.~JP16H00984) from MEXT, and Grant-in-Aid for Scientific Research on Innovative Area ``Nano Spin Conversion Science" (No.~JP15H01012 and No.~JP17H05177) from MEXT.
Part of this research was performed within the Nanometer Structure Consortium/NanoLund-environment, using the facilities of Lund Nano Lab, with support from the Swedish Research Council (VR), the Swedish Foundation for Strategic Research (SSF) and from Knut and Alice Wallenberg Foundation (KAW).

H.K. and R.S.D. contributed equally to this work.

\newpage

\begin{figure}[ht]
\begin{center}
\includegraphics[scale=0.65]{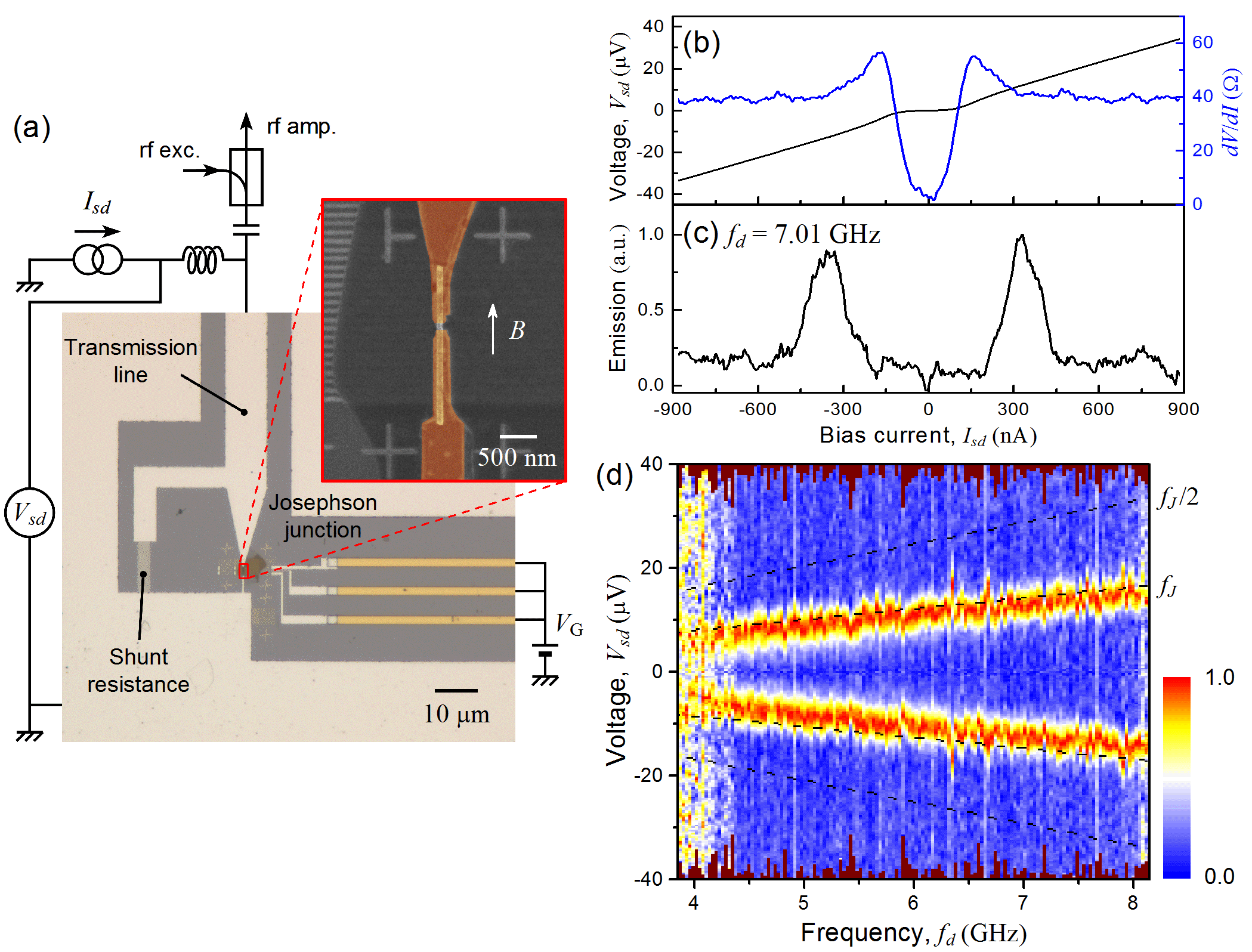}
\caption{
(a)
An optical micrograph of the device and the experimental setup for the emission measurement.
A false-colored scanning electron micrograph (SEM) of the NW Josephson junction is shown in an enlarged view.
On the SEM image, the NW, Al contacts, hBN insulating layer, and local gate electrodes are indicated in white, orange, dark gray, and light gray, respectively.
A shunt resistance made of Cr/Au (8/8~nm) enables a stable voltage bias across the junction.
An on-chip coplanar transmission line (1250~$\mu$m in length and 10~$\mu$m in width) connected to one of the contacts is defined in the same fabrication step as the contacts.
A remote four-terminal measurement setup, in which junction bias is measured between the bias tee and the superconducting ground plane of the device chip, is employed.
Measurements therefore include a series resistance from normal metal measurement components up to the device chip and rely on maintenance of the superconducting state of the transmission line and ground plane.
The transmission line collects the radiation emitted from the device and couples through the bias-tee to the rf measurement setup.
The gate electrodes are biased at $V_{\rm G} = 0$~V in all measurements.
(b) Current-voltage characteristics (black line) and numerically calculated differential resistance $dV/dI$ (blue line) as a function of bias current $I_{sd}$ observed at $B = 0$~mT.
(c) Emission spectrum as a function of bias current $I_{sd}$ observed at zero magnetic field for detection frequency of $f_{d} = 7.01$~GHz.
The data is normalized to maximum emission amplitude.
(d) Emission spectra as a function of voltage bias $V_{sd}$ and detection frequency $f_{d}$ observed at zero magnetic field.
Dashed lines indicate the expected resonance lines of $f_{J}$ (inner) and $f_{J}/2$ (outer).
For better visibility, the data is normalized to the maximum emission amplitude for each $f_{d}$.
}
\label{Fig1}
\end{center}
\end{figure}

\newpage

\begin{figure}[ht]
\begin{center}
\includegraphics[scale=0.75]{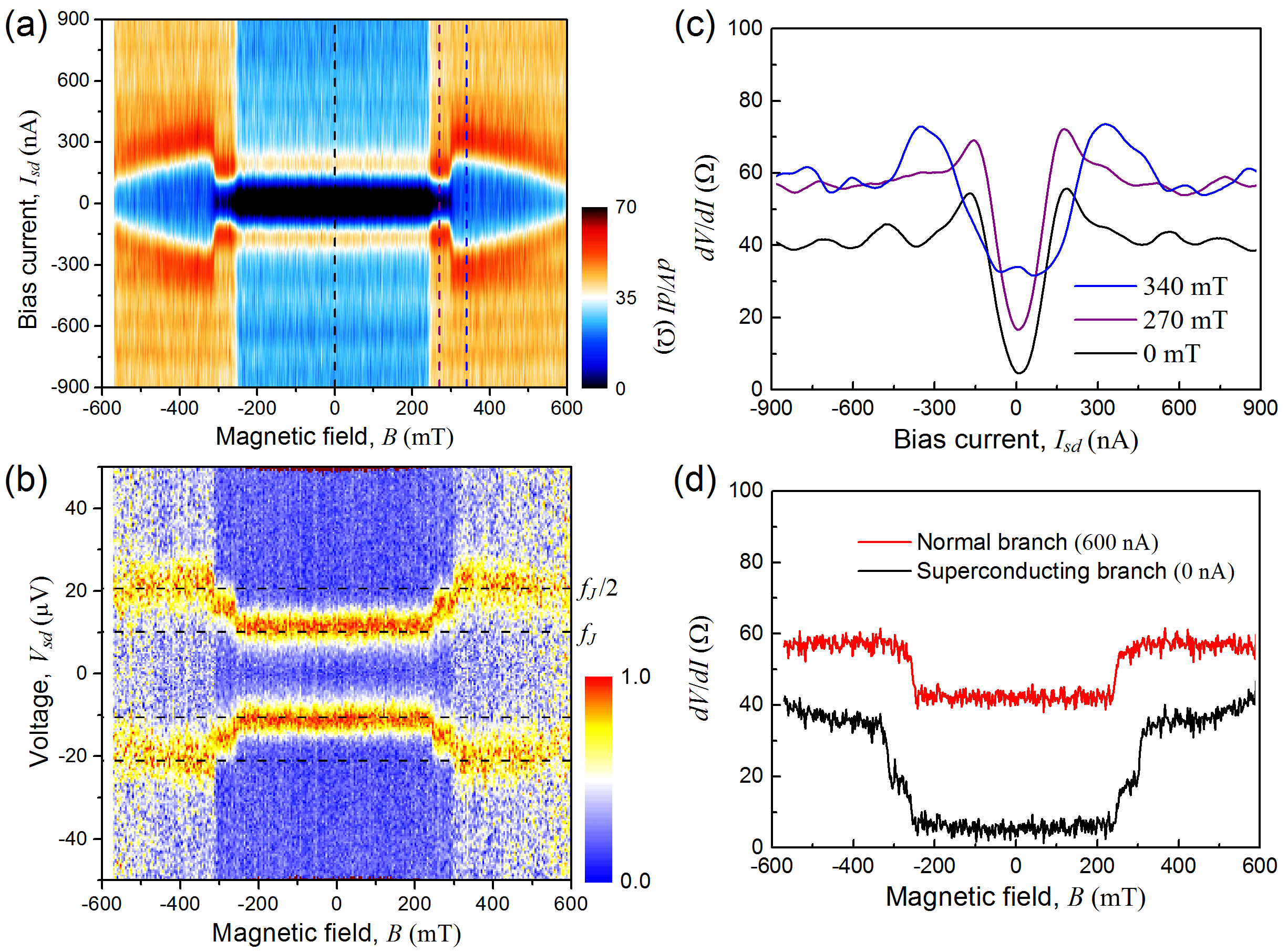}
\caption{
(a)
Differential resistance $dV/dI$ as a function of an in-plane magnetic field $B$ and bias current $I_{sd}$.
(b)
Emission spectra as a function of $B$ and voltage bias $V_{sd}$ observed at fixed detection frequency of $f_{d} = 5.0$~GHz.
Dashed lines indicate the expected resonance lines of $f_{J}$ (inner) and $f_{J} / 2$ (outer).
For better visibility, the data is normalized to the maximum emission amplitude for each $B$.
(c)
Horizontal cuts of (a) at $B = 0$~mT, 270~mT and 340~mT.
(d)
Vertical cuts of (a) in the superconducting ($I_{sd} = 0$~nA) and normal ($I_{sd} = 600$~nA) branches.
}
\label{Fig2}
\end{center}
\end{figure}

\newpage

\begin{figure}[ht]
\begin{center}
\includegraphics[scale=0.8]{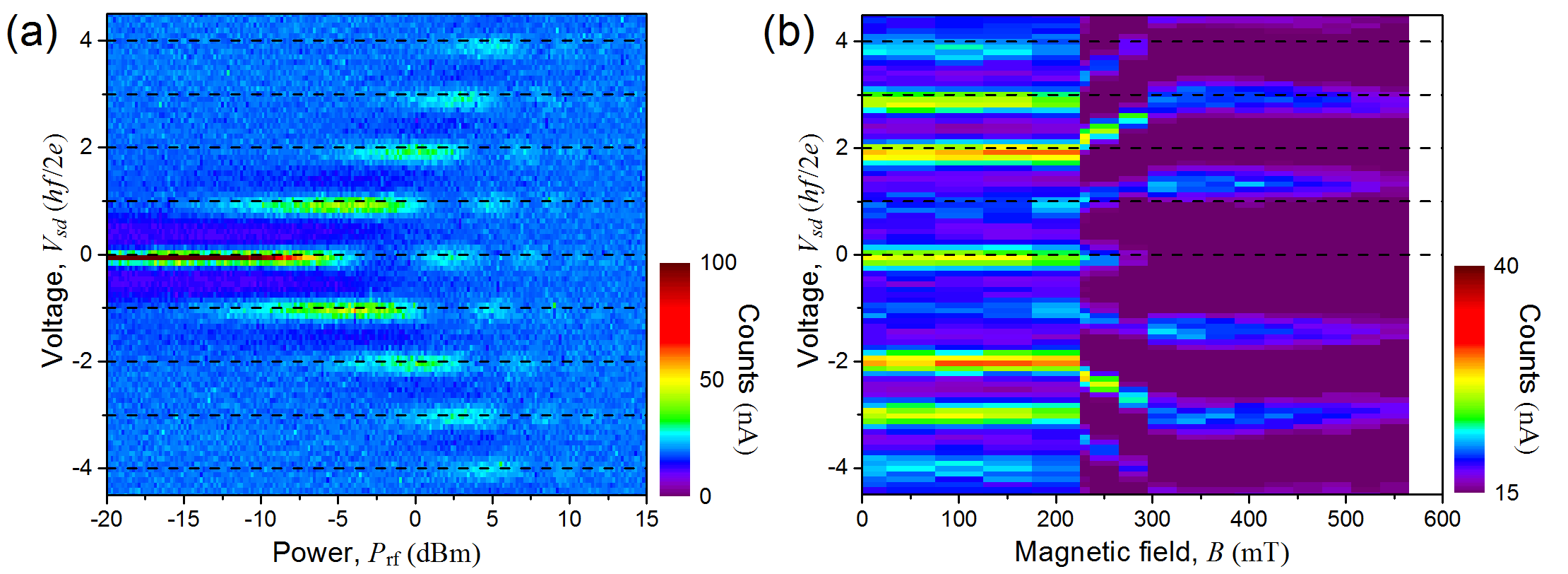}
\caption{
(a)
Two-dimensional color plot of the bin counts of each Shapiro step as a function of voltage bias $V_{sd}$ (in normalized units $hf/2e$) and rf excitation power $P_{\rm rf}$ observed at fixed frequency of $f = 4.0$~GHz and $B = 0$~mT.
(b)
B dependence of the bin counts at fixed frequency of $f = 4.0$~GHz and power of $P_{\rm rf} = 1.0$~dBm.
}
\label{Fig3}
\end{center}
\end{figure}

\newpage

\begin{figure}[ht]
\begin{center}
\includegraphics[scale=0.8]{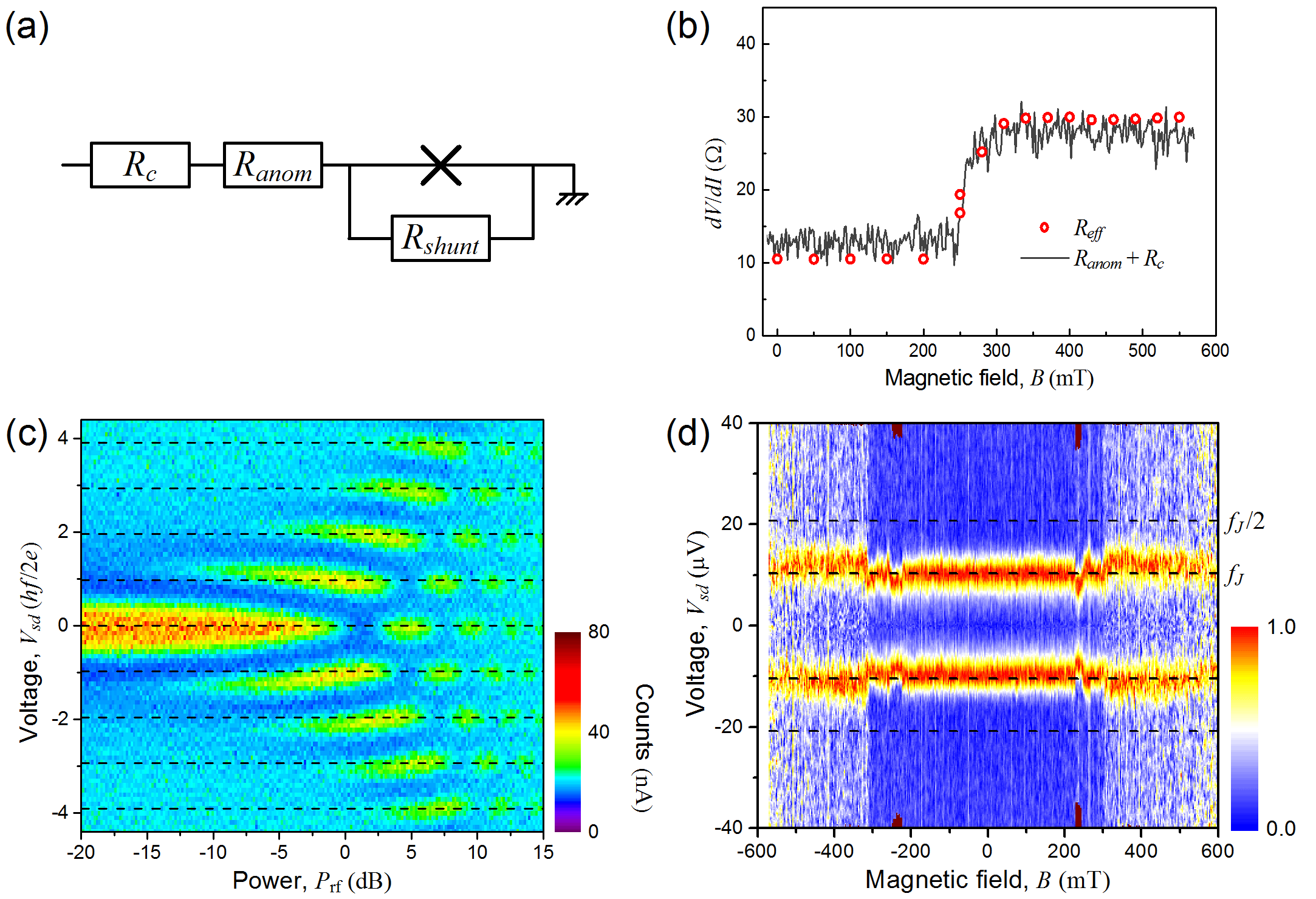}
\caption{
(a)
Schematic circuit diagram of the Josephson junction in series with a constant circuit resistance $R_{c}$ (due to normal external components) and anomalous resistance $R_{anom}$, and in parallel with a shunt resistance $R_{shunt}$.
(b)
Effective resistance $R_{eff}$ (circles) and anomalous resistance $R_{anom}$ plus constant resistance of $R_{c} = 14$~$\Omega$ (solid line) as a function of $B$.
(c)
Corrected Shapiro response observed at fixed frequency of $f = 4.0$~GHz and $B = 340$~mT.
(d)
Corrected emission spectra as a function of $B$ and voltage bias $V_{sd}$ observed at fixed detection frequency of $f_{d} = 5.0$~GHz.
Dashed lines indicate the expected resonance lines of $f_{J}$ (inner) and $f_{J} / 2$ (outer).
For better visibility, the data is normalized to the maximum emission amplitude for each $B$.
}
\label{Fig4}
\end{center}
\end{figure}

\newpage

\renewcommand{\thefigure}{S\arabic{figure}}
\setcounter{figure}{0}

\section*{Supplementary Information}

\noindent{\bf 1. Ballistic transport properties of InAs nanowires}

\begin{figure}[ht]
\begin{center}
\includegraphics[scale=0.8]{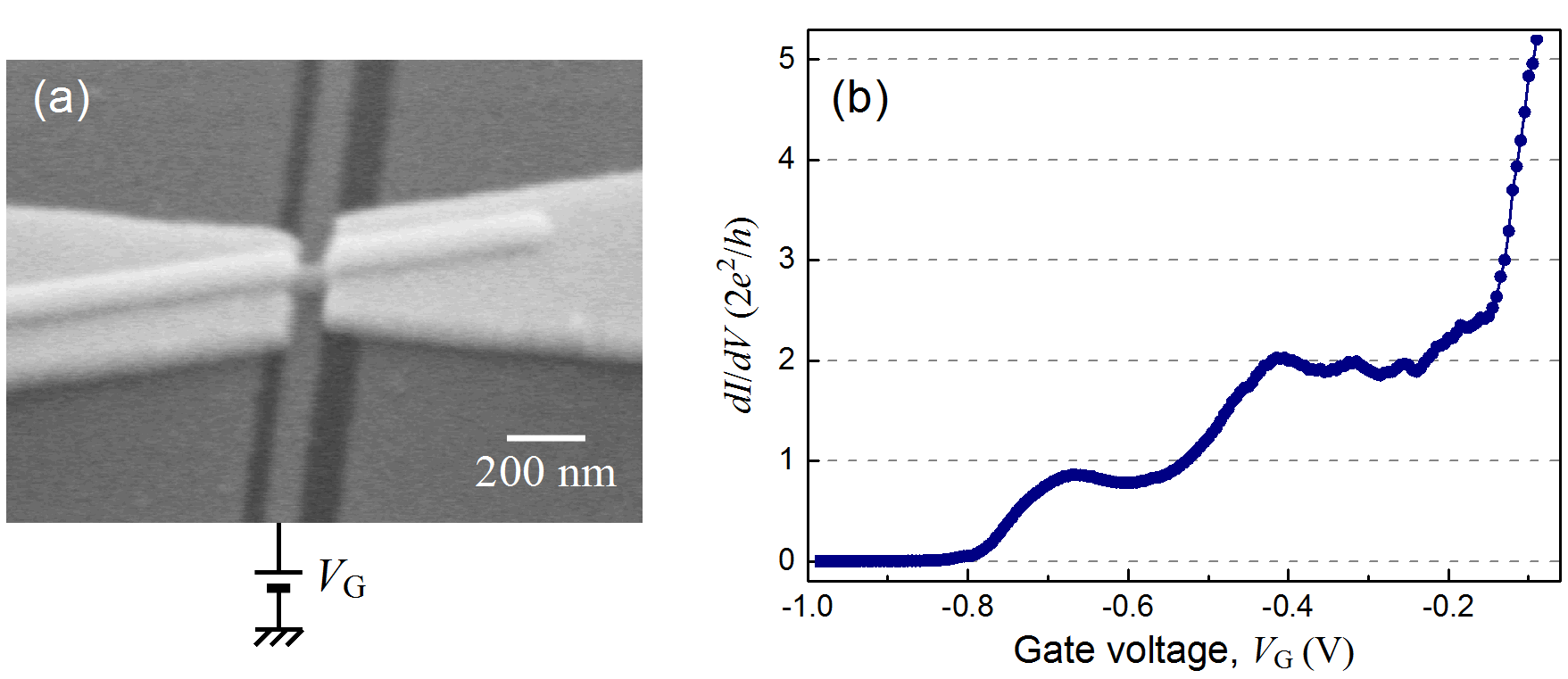}
\caption{
(a)
A scanning electron micrograph of a NW junction with normal contacts.
The spacing of the two contacts is $\sim 100$~nm.
(b)
Zero bias conductance as a function of $V_{\rm G}$ observed at $B = 0$~mT.
}
\label{FigS1}
\end{center}
\end{figure}

It is known that InAs nanowires (NWs) have a charge accumulation layer on the surface \cite{olsson:1996}.
Although ballistic transport in NWs plays a crucial role in creating a topological state in proximitized NWs, conductive electrons in the accumulation layer experience strong surface roughness scattering and ionized impurity scattering such that the transport is entirely diffusive. 
To suppress such scatterings we employ hexagonal boronnitride (hBN) as a gate dielectric.
Figure~\ref{FigS1}(a) shows an InAs NW transferred onto the hBN insulating dielectric with normal contacts made of Ti/Au (3/110~nm).
The conductance of the nanowire is controlled with a voltage $V_{\rm G}$ applied to the gate electrode underneath the junction, and measured with a standard lock-in technique.
Figure~\ref{FigS1}(b) shows the zero bias conductance as a function of $V_{\rm G}$ observed at $B = 0$~mT.
The observed quantized conductance with first and second spin-degenerate subbands provides direct evidence of ballistic transport in the NW.

\noindent{\bf 2. Transport properties of InAs nanowire Josephson junctions with different circuit environments}

\begin{figure}[ht]
\begin{center}
\includegraphics[scale=0.7]{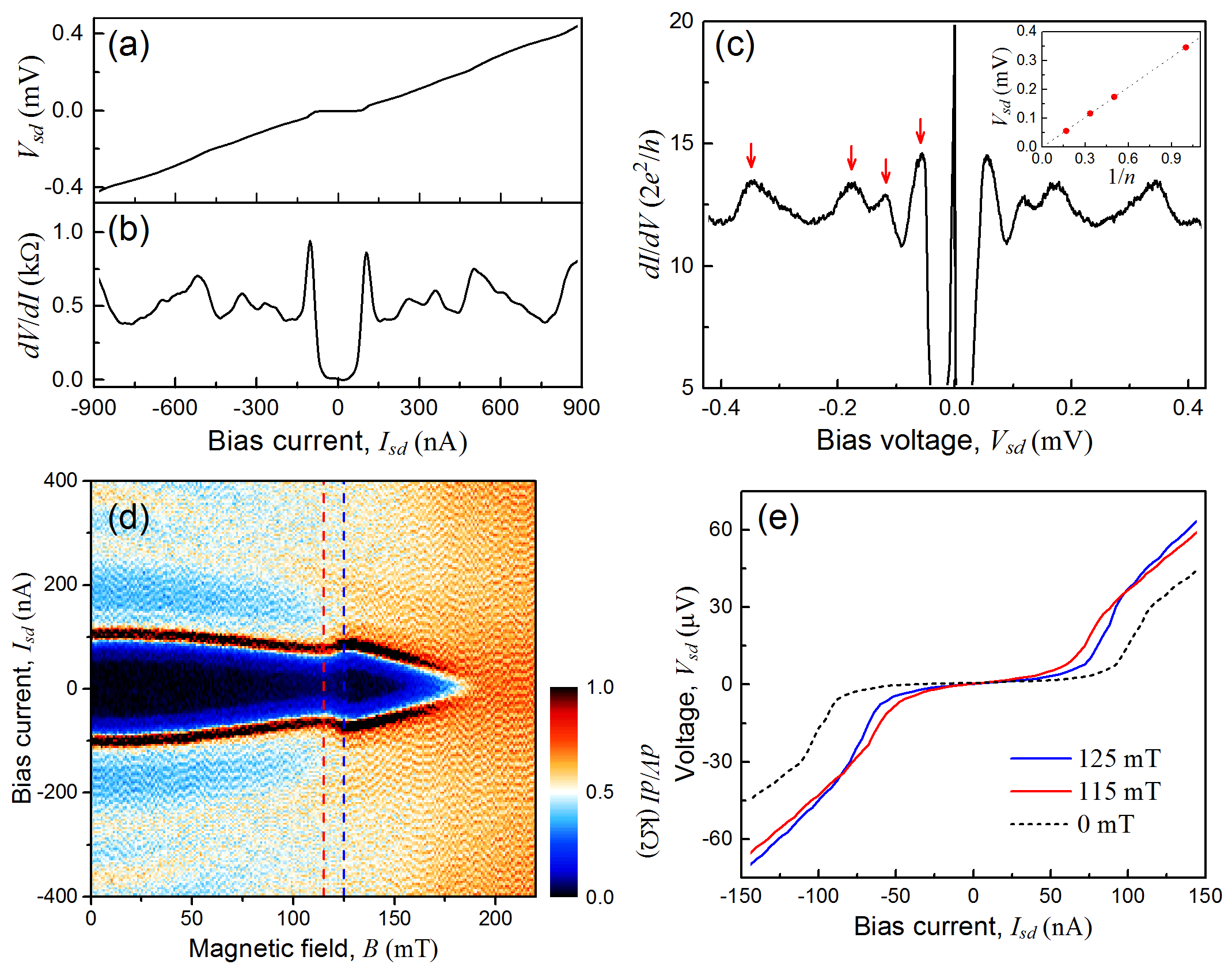}
\caption{
Current-voltage characteristics (a) and numerically calculated differential resistance $dV/dI$ (b) as a function of bias current $I_{sd}$ observed at $B = 0$~mT.
(c) Differential conductance $dI/dV$ as a function of bias voltage $V_{sd}$ observed at $B = 0$~mT.
The pronounced peak observed at $V_{sd} = 0$~V corresponds to a superconducting branch.
The arrows indicate conductance peaks at the voltage positions of $e V_{n} = 2 \Delta^{\ast} / n$~($n = 1, 2, \ldots$), corresponding to MAR processes.
The inset shows the peak position as a function of the inverse peak index $1/n$.
The dashed line indicates a linear fit.
(d) Differential resistance $dV/dI$ as a function of in-plane magnetic field $B$ and bias current $I_{sd}$.
(e) Current-voltage characteristics observed at $B = 0$~mT (black dashed line), 115~mT (red solid line) and 125~mT (blue solid line).
}
\label{FigS2}
\end{center}
\end{figure}

Transport properties of a different InAs NW Josephson junction prepared in the same way without an on-chip transmission line and shunt resistance are summarized in Fig.~\ref{FigS2}.
Compared with the Josephson junction in the main text, this junction has superconducting contacts with thicker Ti/Al (1.5~nm/125~nm) and the same spacing of $\sim 60$~nm.
We measured this junction in another dilution refrigerator, which has different low temperature filters and sample enclosures with Au bonding wires instead of Al.

Figures~\ref{FigS2}(a) and \ref{FigS2}(b) show typical examples of the dc current-voltage characteristic and the numerically calculated differential resistance $dV/dI$ observed at $B = 0$~mT, respectively.
The observed switching current $I_{c} \sim 100$~nA without hysteretic behavior is consistent with that shown in the main text.
We note that some ripples observed in the normal state are related to self-induced Shapiro steps \cite{dayem:1966}.

We evaluate an induced superconducting gap $\Delta^{\ast}$ in this junction by measuring multiple Andreev reflection (MAR) resonances.
Figure~\ref{FigS2}(c) shows differential conductance $dI/dV$ as a function of bias voltage $V_{sd}$ observed at $B = 0$~mT.
The observed $dI/dV$ displays pronounced MAR features, showing sharp peaks as indicated by arrows.
From the MAR relation $e V_{n} = 2 \Delta^{\ast} / n$~($n = 1, 2, \ldots$), a linear fit of the inverse peak index $1/n$ as a function of bias voltage $V_{sd}$ [inset of Fig.~\ref{FigS2}(c)] yields the induced gap $\Delta^{\ast} \sim 0.18$~meV, which is close to the value for bulk Al.

Importantly, this junction also shows an enhancement of the switching current $I_{c}$ in the finite magnetic field regime as seen in Fig.~2(a) of the main text.
Figure~\ref{FigS2}(d) shows differential resistance $dV/dI$ as a function of in-plane magnetic field $B$ and bias current $I_{sd}$.
As $B$ is increased, while the switching current $I_{c}$ monotonically decreases, $I_{c}$ is enhanced at $B \sim 120$~mT (see red and blue solid lines in Fig.~\ref{FigS2}(e)), and decays at larger $B$.
Although some differences in experimental setups lead to quantitative differences in the critical field and the amplitude of the $I_{c}$ enhancement, this behavior is qualitatively consistent with the trend we observe in Fig.~2(a) of the main text.

\newpage

\newpage

\begin{thebibliography}{31}%
\makeatletter
\providecommand \@ifxundefined [1]{%
 \@ifx{#1\undefined}
}%
\providecommand \@ifnum [1]{%
 \ifnum #1\expandafter \@firstoftwo
 \else \expandafter \@secondoftwo
 \fi
}%
\providecommand \@ifx [1]{%
 \ifx #1\expandafter \@firstoftwo
 \else \expandafter \@secondoftwo
 \fi
}%
\providecommand \natexlab [1]{#1}%
\providecommand \enquote  [1]{``#1''}%
\providecommand \bibnamefont  [1]{#1}%
\providecommand \bibfnamefont [1]{#1}%
\providecommand \citenamefont [1]{#1}%
\providecommand \href@noop [0]{\@secondoftwo}%
\providecommand \href [0]{\begingroup \@sanitize@url \@href}%
\providecommand \@href[1]{\@@startlink{#1}\@@href}%
\providecommand \@@href[1]{\endgroup#1\@@endlink}%
\providecommand \@sanitize@url [0]{\catcode `\\12\catcode `\$12\catcode
  `\&12\catcode `\#12\catcode `\^12\catcode `\_12\catcode `\%12\relax}%
\providecommand \@@startlink[1]{}%
\providecommand \@@endlink[0]{}%
\providecommand \url  [0]{\begingroup\@sanitize@url \@url }%
\providecommand \@url [1]{\endgroup\@href {#1}{\urlprefix }}%
\providecommand \urlprefix  [0]{URL }%
\providecommand \Eprint [0]{\href }%
\providecommand \doibase [0]{http://dx.doi.org/}%
\providecommand \selectlanguage [0]{\@gobble}%
\providecommand \bibinfo  [0]{\@secondoftwo}%
\providecommand \bibfield  [0]{\@secondoftwo}%
\providecommand \translation [1]{[#1]}%
\providecommand \BibitemOpen [0]{}%
\providecommand \bibitemStop [0]{}%
\providecommand \bibitemNoStop [0]{.\EOS\space}%
\providecommand \EOS [0]{\spacefactor3000\relax}%
\providecommand \BibitemShut  [1]{\csname bibitem#1\endcsname}%
\let\auto@bib@innerbib\@empty
\bibitem [{\citenamefont {Nayak}\ \emph {et~al.}(2008)\citenamefont {Nayak},
  \citenamefont {Simon}, \citenamefont {Stern}, \citenamefont {Freedman},\ and\
  \citenamefont {Das~Sarma}}]{nayak:2008}%
  \BibitemOpen
  \bibfield  {author} {\bibinfo {author} {\bibfnamefont {C.}~\bibnamefont
  {Nayak}}, \bibinfo {author} {\bibfnamefont {S.~H.}\ \bibnamefont {Simon}},
  \bibinfo {author} {\bibfnamefont {A.}~\bibnamefont {Stern}}, \bibinfo
  {author} {\bibfnamefont {M.}~\bibnamefont {Freedman}}, \ and\ \bibinfo
  {author} {\bibfnamefont {S.}~\bibnamefont {Das~Sarma}},\ }\href {\doibase
  10.1103/RevModPhys.80.1083} {\bibfield  {journal} {\bibinfo  {journal} {Rev.
  Mod. Phys.}\ }\textbf {\bibinfo {volume} {80}},\ \bibinfo {pages} {1083}
  (\bibinfo {year} {2008})}\BibitemShut {NoStop}%
\bibitem [{\citenamefont {Kitaev}(2001)}]{kitaev:2001}%
  \BibitemOpen
  \bibfield  {author} {\bibinfo {author} {\bibfnamefont {A.~Y.}\ \bibnamefont
  {Kitaev}},\ }\href {http://stacks.iop.org/1063-7869/44/i=10S/a=S29}
  {\bibfield  {journal} {\bibinfo  {journal} {Physics-Uspekhi}\ }\textbf
  {\bibinfo {volume} {44}},\ \bibinfo {pages} {131} (\bibinfo {year}
  {2001})}\BibitemShut {NoStop}%
\bibitem [{\citenamefont {Lutchyn}\ \emph {et~al.}(2010)\citenamefont
  {Lutchyn}, \citenamefont {Sau},\ and\ \citenamefont
  {Das~Sarma}}]{lutchyn:2010}%
  \BibitemOpen
  \bibfield  {author} {\bibinfo {author} {\bibfnamefont {R.~M.}\ \bibnamefont
  {Lutchyn}}, \bibinfo {author} {\bibfnamefont {J.~D.}\ \bibnamefont {Sau}}, \
  and\ \bibinfo {author} {\bibfnamefont {S.}~\bibnamefont {Das~Sarma}},\ }\href
  {\doibase 10.1103/PhysRevLett.105.077001} {\bibfield  {journal} {\bibinfo
  {journal} {Phys. Rev. Lett.}\ }\textbf {\bibinfo {volume} {105}},\ \bibinfo
  {pages} {077001} (\bibinfo {year} {2010})}\BibitemShut {NoStop}%
\bibitem [{\citenamefont {Oreg}\ \emph {et~al.}(2010)\citenamefont {Oreg},
  \citenamefont {Refael},\ and\ \citenamefont {von Oppen}}]{oreg:2010}%
  \BibitemOpen
  \bibfield  {author} {\bibinfo {author} {\bibfnamefont {Y.}~\bibnamefont
  {Oreg}}, \bibinfo {author} {\bibfnamefont {G.}~\bibnamefont {Refael}}, \ and\
  \bibinfo {author} {\bibfnamefont {F.}~\bibnamefont {von Oppen}},\ }\href
  {\doibase 10.1103/PhysRevLett.105.177002} {\bibfield  {journal} {\bibinfo
  {journal} {Phys. Rev. Lett.}\ }\textbf {\bibinfo {volume} {105}},\ \bibinfo
  {pages} {177002} (\bibinfo {year} {2010})}\BibitemShut {NoStop}%
\bibitem [{\citenamefont {Mourik}\ \emph {et~al.}(2012)\citenamefont {Mourik},
  \citenamefont {Zuo}, \citenamefont {Frolov}, \citenamefont {Plissard},
  \citenamefont {Bakkers},\ and\ \citenamefont {Kouwenhoven}}]{mourik:2012}%
  \BibitemOpen
  \bibfield  {author} {\bibinfo {author} {\bibfnamefont {V.}~\bibnamefont
  {Mourik}}, \bibinfo {author} {\bibfnamefont {K.}~\bibnamefont {Zuo}},
  \bibinfo {author} {\bibfnamefont {S.~M.}\ \bibnamefont {Frolov}}, \bibinfo
  {author} {\bibfnamefont {S.~R.}\ \bibnamefont {Plissard}}, \bibinfo {author}
  {\bibfnamefont {E.~P. A.~M.}\ \bibnamefont {Bakkers}}, \ and\ \bibinfo
  {author} {\bibfnamefont {L.~P.}\ \bibnamefont {Kouwenhoven}},\ }\href
  {\doibase 10.1126/science.1222360} {\bibfield  {journal} {\bibinfo  {journal}
  {Science}\ }\textbf {\bibinfo {volume} {336}},\ \bibinfo {pages} {1003}
  (\bibinfo {year} {2012})}\BibitemShut {NoStop}%
\bibitem [{\citenamefont {Das}\ \emph {et~al.}(2012)\citenamefont {Das},
  \citenamefont {Ronen}, \citenamefont {Most}, \citenamefont {Oreg},
  \citenamefont {Heiblum},\ and\ \citenamefont {Shtrikman}}]{das:2012}%
  \BibitemOpen
  \bibfield  {author} {\bibinfo {author} {\bibfnamefont {A.}~\bibnamefont
  {Das}}, \bibinfo {author} {\bibfnamefont {Y.}~\bibnamefont {Ronen}}, \bibinfo
  {author} {\bibfnamefont {Y.}~\bibnamefont {Most}}, \bibinfo {author}
  {\bibfnamefont {Y.}~\bibnamefont {Oreg}}, \bibinfo {author} {\bibfnamefont
  {M.}~\bibnamefont {Heiblum}}, \ and\ \bibinfo {author} {\bibfnamefont
  {H.}~\bibnamefont {Shtrikman}},\ }\href {http://dx.doi.org/10.1038/nphys2479}
  {\bibfield  {journal} {\bibinfo  {journal} {Nature Physics}\ }\textbf
  {\bibinfo {volume} {8}},\ \bibinfo {pages} {887} (\bibinfo {year}
  {2012})}\BibitemShut {NoStop}%
\bibitem [{\citenamefont {Deng}\ \emph {et~al.}(2016)\citenamefont {Deng},
  \citenamefont {Vaitiekenas}, \citenamefont {Hansen}, \citenamefont {Danon},
  \citenamefont {Leijnse}, \citenamefont {Flensberg}, \citenamefont {Nyg{\r
  a}rd}, \citenamefont {Krogstrup},\ and\ \citenamefont {Marcus}}]{deng:2016}%
  \BibitemOpen
  \bibfield  {author} {\bibinfo {author} {\bibfnamefont {M.~T.}\ \bibnamefont
  {Deng}}, \bibinfo {author} {\bibfnamefont {S.}~\bibnamefont {Vaitiekenas}},
  \bibinfo {author} {\bibfnamefont {E.~B.}\ \bibnamefont {Hansen}}, \bibinfo
  {author} {\bibfnamefont {J.}~\bibnamefont {Danon}}, \bibinfo {author}
  {\bibfnamefont {M.}~\bibnamefont {Leijnse}}, \bibinfo {author} {\bibfnamefont
  {K.}~\bibnamefont {Flensberg}}, \bibinfo {author} {\bibfnamefont
  {J.}~\bibnamefont {Nyg{\r a}rd}}, \bibinfo {author} {\bibfnamefont
  {P.}~\bibnamefont {Krogstrup}}, \ and\ \bibinfo {author} {\bibfnamefont
  {C.~M.}\ \bibnamefont {Marcus}},\ }\href {\doibase 10.1126/science.aaf3961}
  {\bibfield  {journal} {\bibinfo  {journal} {Science}\ }\textbf {\bibinfo
  {volume} {354}},\ \bibinfo {pages} {1557} (\bibinfo {year}
  {2016})}\BibitemShut {NoStop}%
\bibitem [{\citenamefont {Zhang}\ \emph {et~al.}(2017)\citenamefont {Zhang},
  \citenamefont {Liu}, \citenamefont {Gazibegovic}, \citenamefont {Xu},
  \citenamefont {Logan}, \citenamefont {Wang}, \citenamefont {van Loo},
  \citenamefont {Bommer}, \citenamefont {de~Moor}, \citenamefont {Car},
  \citenamefont {het Veld}, \citenamefont {van Veldhoven}, \citenamefont
  {Koelling}, \citenamefont {Verheijen}, \citenamefont {Pendharkar},
  \citenamefont {Pennachio}, \citenamefont {Shojaei}, \citenamefont {Lee},
  \citenamefont {Palmstrom}, \citenamefont {Bakkers}, \citenamefont {Sarma},\
  and\ \citenamefont {Kouwenhoven}}]{zhang:2017}%
  \BibitemOpen
  \bibfield  {author} {\bibinfo {author} {\bibfnamefont {H.}~\bibnamefont
  {Zhang}}, \bibinfo {author} {\bibfnamefont {C.-X.}\ \bibnamefont {Liu}},
  \bibinfo {author} {\bibfnamefont {S.}~\bibnamefont {Gazibegovic}}, \bibinfo
  {author} {\bibfnamefont {D.}~\bibnamefont {Xu}}, \bibinfo {author}
  {\bibfnamefont {J.~A.}\ \bibnamefont {Logan}}, \bibinfo {author}
  {\bibfnamefont {G.}~\bibnamefont {Wang}}, \bibinfo {author} {\bibfnamefont
  {N.}~\bibnamefont {van Loo}}, \bibinfo {author} {\bibfnamefont {J.~D.~S.}\
  \bibnamefont {Bommer}}, \bibinfo {author} {\bibfnamefont {M.~W.~A.}\
  \bibnamefont {de~Moor}}, \bibinfo {author} {\bibfnamefont {D.}~\bibnamefont
  {Car}}, \bibinfo {author} {\bibfnamefont {R.~L. M.~O.}\ \bibnamefont {het
  Veld}}, \bibinfo {author} {\bibfnamefont {P.~J.}\ \bibnamefont {van
  Veldhoven}}, \bibinfo {author} {\bibfnamefont {S.}~\bibnamefont {Koelling}},
  \bibinfo {author} {\bibfnamefont {M.~A.}\ \bibnamefont {Verheijen}}, \bibinfo
  {author} {\bibfnamefont {M.}~\bibnamefont {Pendharkar}}, \bibinfo {author}
  {\bibfnamefont {D.~J.}\ \bibnamefont {Pennachio}}, \bibinfo {author}
  {\bibfnamefont {B.}~\bibnamefont {Shojaei}}, \bibinfo {author} {\bibfnamefont
  {J.~S.}\ \bibnamefont {Lee}}, \bibinfo {author} {\bibfnamefont {C.~J.}\
  \bibnamefont {Palmstrom}}, \bibinfo {author} {\bibfnamefont {E.~P. A.~M.}\
  \bibnamefont {Bakkers}}, \bibinfo {author} {\bibfnamefont {S.~D.}\
  \bibnamefont {Sarma}}, \ and\ \bibinfo {author} {\bibfnamefont {L.~P.}\
  \bibnamefont {Kouwenhoven}},\ }\href@noop {} {\enquote {\bibinfo {title}
  {Quantized majorana conductance},}\ } (\bibinfo {year} {2017}),\ \Eprint
  {http://arxiv.org/abs/arXiv:1710.10701} {arXiv:1710.10701} \BibitemShut
  {NoStop}%
\bibitem [{\citenamefont {Nichele}\ \emph {et~al.}(2017)\citenamefont
  {Nichele}, \citenamefont {Drachmann}, \citenamefont {Whiticar}, \citenamefont
  {O'Farrell}, \citenamefont {Suominen}, \citenamefont {Fornieri},
  \citenamefont {Wang}, \citenamefont {Gardner}, \citenamefont {Thomas},
  \citenamefont {Hatke}, \citenamefont {Krogstrup}, \citenamefont {Manfra},
  \citenamefont {Flensberg},\ and\ \citenamefont {Marcus}}]{nichele:2017}%
  \BibitemOpen
  \bibfield  {author} {\bibinfo {author} {\bibfnamefont {F.}~\bibnamefont
  {Nichele}}, \bibinfo {author} {\bibfnamefont {A.~C.~C.}\ \bibnamefont
  {Drachmann}}, \bibinfo {author} {\bibfnamefont {A.~M.}\ \bibnamefont
  {Whiticar}}, \bibinfo {author} {\bibfnamefont {E.~C.~T.}\ \bibnamefont
  {O'Farrell}}, \bibinfo {author} {\bibfnamefont {H.~J.}\ \bibnamefont
  {Suominen}}, \bibinfo {author} {\bibfnamefont {A.}~\bibnamefont {Fornieri}},
  \bibinfo {author} {\bibfnamefont {T.}~\bibnamefont {Wang}}, \bibinfo {author}
  {\bibfnamefont {G.~C.}\ \bibnamefont {Gardner}}, \bibinfo {author}
  {\bibfnamefont {C.}~\bibnamefont {Thomas}}, \bibinfo {author} {\bibfnamefont
  {A.~T.}\ \bibnamefont {Hatke}}, \bibinfo {author} {\bibfnamefont
  {P.}~\bibnamefont {Krogstrup}}, \bibinfo {author} {\bibfnamefont {M.~J.}\
  \bibnamefont {Manfra}}, \bibinfo {author} {\bibfnamefont {K.}~\bibnamefont
  {Flensberg}}, \ and\ \bibinfo {author} {\bibfnamefont {C.~M.}\ \bibnamefont
  {Marcus}},\ }\href {\doibase 10.1103/PhysRevLett.119.136803} {\bibfield
  {journal} {\bibinfo  {journal} {Phys. Rev. Lett.}\ }\textbf {\bibinfo
  {volume} {119}},\ \bibinfo {pages} {136803} (\bibinfo {year}
  {2017})}\BibitemShut {NoStop}%
\bibitem [{\citenamefont {Suominen}\ \emph {et~al.}(2017)\citenamefont
  {Suominen}, \citenamefont {Kjaergaard}, \citenamefont {Hamilton},
  \citenamefont {Shabani}, \citenamefont {Palmstr\o{}m}, \citenamefont
  {Marcus},\ and\ \citenamefont {Nichele}}]{suominen:2017}%
  \BibitemOpen
  \bibfield  {author} {\bibinfo {author} {\bibfnamefont {H.~J.}\ \bibnamefont
  {Suominen}}, \bibinfo {author} {\bibfnamefont {M.}~\bibnamefont
  {Kjaergaard}}, \bibinfo {author} {\bibfnamefont {A.~R.}\ \bibnamefont
  {Hamilton}}, \bibinfo {author} {\bibfnamefont {J.}~\bibnamefont {Shabani}},
  \bibinfo {author} {\bibfnamefont {C.~J.}\ \bibnamefont {Palmstr\o{}m}},
  \bibinfo {author} {\bibfnamefont {C.~M.}\ \bibnamefont {Marcus}}, \ and\
  \bibinfo {author} {\bibfnamefont {F.}~\bibnamefont {Nichele}},\ }\href
  {\doibase 10.1103/PhysRevLett.119.176805} {\bibfield  {journal} {\bibinfo
  {journal} {Phys. Rev. Lett.}\ }\textbf {\bibinfo {volume} {119}},\ \bibinfo
  {pages} {176805} (\bibinfo {year} {2017})}\BibitemShut {NoStop}%
\bibitem [{\citenamefont {Beenakker}(2013)}]{beenakker:2013}%
  \BibitemOpen
  \bibfield  {author} {\bibinfo {author} {\bibfnamefont {C.}~\bibnamefont
  {Beenakker}},\ }\href {\doibase 10.1146/annurev-conmatphys-030212-184337}
  {\bibfield  {journal} {\bibinfo  {journal} {Annual Review of Condensed Matter
  Physics}\ }\textbf {\bibinfo {volume} {4}},\ \bibinfo {pages} {113} (\bibinfo
  {year} {2013})}\BibitemShut {NoStop}%
\bibitem [{\citenamefont {Kwon}\ \emph {et~al.}(2004)\citenamefont {Kwon},
  \citenamefont {Sengupta},\ and\ \citenamefont {Yakovenko}}]{kwon:2004}%
  \BibitemOpen
  \bibfield  {author} {\bibinfo {author} {\bibfnamefont {H.-J.}\ \bibnamefont
  {Kwon}}, \bibinfo {author} {\bibfnamefont {K.}~\bibnamefont {Sengupta}}, \
  and\ \bibinfo {author} {\bibfnamefont {V.~M.}\ \bibnamefont {Yakovenko}},\
  }\href {\doibase 10.1140/epjb/e2004-00066-4} {\bibfield  {journal} {\bibinfo
  {journal} {The European Physical Journal B - Condensed Matter and Complex
  Systems}\ }\textbf {\bibinfo {volume} {37}},\ \bibinfo {pages} {349}
  (\bibinfo {year} {2004})}\BibitemShut {NoStop}%
\bibitem [{\citenamefont {Tiira}\ \emph {et~al.}(2017)\citenamefont {Tiira},
  \citenamefont {Strambini}, \citenamefont {Amado}, \citenamefont {Roddaro},
  \citenamefont {San-Jose}, \citenamefont {Aguado}, \citenamefont {Bergeret},
  \citenamefont {Ercolani}, \citenamefont {Sorba},\ and\ \citenamefont
  {Giazotto}}]{tiira:2017}%
  \BibitemOpen
  \bibfield  {author} {\bibinfo {author} {\bibfnamefont {J.}~\bibnamefont
  {Tiira}}, \bibinfo {author} {\bibfnamefont {E.}~\bibnamefont {Strambini}},
  \bibinfo {author} {\bibfnamefont {M.}~\bibnamefont {Amado}}, \bibinfo
  {author} {\bibfnamefont {S.}~\bibnamefont {Roddaro}}, \bibinfo {author}
  {\bibfnamefont {P.}~\bibnamefont {San-Jose}}, \bibinfo {author}
  {\bibfnamefont {R.}~\bibnamefont {Aguado}}, \bibinfo {author} {\bibfnamefont
  {F.~S.}\ \bibnamefont {Bergeret}}, \bibinfo {author} {\bibfnamefont
  {D.}~\bibnamefont {Ercolani}}, \bibinfo {author} {\bibfnamefont
  {L.}~\bibnamefont {Sorba}}, \ and\ \bibinfo {author} {\bibfnamefont
  {F.}~\bibnamefont {Giazotto}},\ }\href
  {http://dx.doi.org/10.1038/ncomms14984} {\bibfield  {journal} {\bibinfo
  {journal} {Nature Communications}\ }\textbf {\bibinfo {volume} {8}},\
  \bibinfo {pages} {14984} (\bibinfo {year} {2017})}\BibitemShut {NoStop}%
\bibitem [{\citenamefont {Rokhinson}\ \emph {et~al.}(2012)\citenamefont
  {Rokhinson}, \citenamefont {Liu},\ and\ \citenamefont
  {Furdyna}}]{rokhinson:2012}%
  \BibitemOpen
  \bibfield  {author} {\bibinfo {author} {\bibfnamefont {L.~P.}\ \bibnamefont
  {Rokhinson}}, \bibinfo {author} {\bibfnamefont {X.}~\bibnamefont {Liu}}, \
  and\ \bibinfo {author} {\bibfnamefont {J.~K.}\ \bibnamefont {Furdyna}},\
  }\href {http://dx.doi.org/10.1038/nphys2429} {\bibfield  {journal} {\bibinfo
  {journal} {Nature Physics}\ }\textbf {\bibinfo {volume} {8}},\ \bibinfo
  {pages} {795} (\bibinfo {year} {2012})}\BibitemShut {NoStop}%
\bibitem [{\citenamefont {Laroche}\ \emph {et~al.}(2017)\citenamefont
  {Laroche}, \citenamefont {Bouman}, \citenamefont {van Woerkom}, \citenamefont
  {Proutski}, \citenamefont {Murthy}, \citenamefont {Pikulin}, \citenamefont
  {Nayak}, \citenamefont {van Gulik}, \citenamefont {Nyg\r{a}rd}, \citenamefont
  {Krogstrup}, \citenamefont {Kouwenhoven},\ and\ \citenamefont
  {Geresdi}}]{laroche:2017}%
  \BibitemOpen
  \bibfield  {author} {\bibinfo {author} {\bibfnamefont {D.}~\bibnamefont
  {Laroche}}, \bibinfo {author} {\bibfnamefont {D.}~\bibnamefont {Bouman}},
  \bibinfo {author} {\bibfnamefont {D.~J.}\ \bibnamefont {van Woerkom}},
  \bibinfo {author} {\bibfnamefont {A.}~\bibnamefont {Proutski}}, \bibinfo
  {author} {\bibfnamefont {C.}~\bibnamefont {Murthy}}, \bibinfo {author}
  {\bibfnamefont {D.~I.}\ \bibnamefont {Pikulin}}, \bibinfo {author}
  {\bibfnamefont {C.}~\bibnamefont {Nayak}}, \bibinfo {author} {\bibfnamefont
  {R.~J.~J.}\ \bibnamefont {van Gulik}}, \bibinfo {author} {\bibfnamefont
  {J.}~\bibnamefont {Nyg\r{a}rd}}, \bibinfo {author} {\bibfnamefont
  {P.}~\bibnamefont {Krogstrup}}, \bibinfo {author} {\bibfnamefont {L.~P.}\
  \bibnamefont {Kouwenhoven}}, \ and\ \bibinfo {author} {\bibfnamefont
  {A.}~\bibnamefont {Geresdi}},\ }\href@noop {} {\enquote {\bibinfo {title}
  {Observation of the 4$\pi$-periodic {J}osephson effect in {I}n{A}s
  nanowires},}\ } (\bibinfo {year} {2017}),\ \Eprint
  {http://arxiv.org/abs/arXiv:1712.08459} {arXiv:1712.08459} \BibitemShut
  {NoStop}%
\bibitem [{\citenamefont {K{\"o}nig}\ \emph {et~al.}(2007)\citenamefont
  {K{\"o}nig}, \citenamefont {Wiedmann}, \citenamefont {Br{\"u}ne},
  \citenamefont {Roth}, \citenamefont {Buhmann}, \citenamefont {Molenkamp},
  \citenamefont {Qi},\ and\ \citenamefont {Zhang}}]{konig:2007}%
  \BibitemOpen
  \bibfield  {author} {\bibinfo {author} {\bibfnamefont {M.}~\bibnamefont
  {K{\"o}nig}}, \bibinfo {author} {\bibfnamefont {S.}~\bibnamefont {Wiedmann}},
  \bibinfo {author} {\bibfnamefont {C.}~\bibnamefont {Br{\"u}ne}}, \bibinfo
  {author} {\bibfnamefont {A.}~\bibnamefont {Roth}}, \bibinfo {author}
  {\bibfnamefont {H.}~\bibnamefont {Buhmann}}, \bibinfo {author} {\bibfnamefont
  {L.~W.}\ \bibnamefont {Molenkamp}}, \bibinfo {author} {\bibfnamefont {X.-L.}\
  \bibnamefont {Qi}}, \ and\ \bibinfo {author} {\bibfnamefont {S.-C.}\
  \bibnamefont {Zhang}},\ }\href {\doibase 10.1126/science.1148047} {\bibfield
  {journal} {\bibinfo  {journal} {Science}\ }\textbf {\bibinfo {volume}
  {318}},\ \bibinfo {pages} {766} (\bibinfo {year} {2007})}\BibitemShut
  {NoStop}%
\bibitem [{\citenamefont {Br\"une}\ \emph {et~al.}(2011)\citenamefont
  {Br\"une}, \citenamefont {Liu}, \citenamefont {Novik}, \citenamefont
  {Hankiewicz}, \citenamefont {Buhmann}, \citenamefont {Chen}, \citenamefont
  {Qi}, \citenamefont {Shen}, \citenamefont {Zhang},\ and\ \citenamefont
  {Molenkamp}}]{brune:2011}%
  \BibitemOpen
  \bibfield  {author} {\bibinfo {author} {\bibfnamefont {C.}~\bibnamefont
  {Br\"une}}, \bibinfo {author} {\bibfnamefont {C.~X.}\ \bibnamefont {Liu}},
  \bibinfo {author} {\bibfnamefont {E.~G.}\ \bibnamefont {Novik}}, \bibinfo
  {author} {\bibfnamefont {E.~M.}\ \bibnamefont {Hankiewicz}}, \bibinfo
  {author} {\bibfnamefont {H.}~\bibnamefont {Buhmann}}, \bibinfo {author}
  {\bibfnamefont {Y.~L.}\ \bibnamefont {Chen}}, \bibinfo {author}
  {\bibfnamefont {X.~L.}\ \bibnamefont {Qi}}, \bibinfo {author} {\bibfnamefont
  {Z.~X.}\ \bibnamefont {Shen}}, \bibinfo {author} {\bibfnamefont {S.~C.}\
  \bibnamefont {Zhang}}, \ and\ \bibinfo {author} {\bibfnamefont {L.~W.}\
  \bibnamefont {Molenkamp}},\ }\href {\doibase 10.1103/PhysRevLett.106.126803}
  {\bibfield  {journal} {\bibinfo  {journal} {Phys. Rev. Lett.}\ }\textbf
  {\bibinfo {volume} {106}},\ \bibinfo {pages} {126803} (\bibinfo {year}
  {2011})}\BibitemShut {NoStop}%
\bibitem [{\citenamefont {Deacon}\ \emph {et~al.}(2017)\citenamefont {Deacon},
  \citenamefont {Wiedenmann}, \citenamefont {Bocquillon}, \citenamefont
  {Dom\'{\i}nguez}, \citenamefont {Klapwijk}, \citenamefont {Leubner},
  \citenamefont {Br\"une}, \citenamefont {Hankiewicz}, \citenamefont {Tarucha},
  \citenamefont {Ishibashi}, \citenamefont {Buhmann},\ and\ \citenamefont
  {Molenkamp}}]{deacon:2017}%
  \BibitemOpen
  \bibfield  {author} {\bibinfo {author} {\bibfnamefont {R.~S.}\ \bibnamefont
  {Deacon}}, \bibinfo {author} {\bibfnamefont {J.}~\bibnamefont {Wiedenmann}},
  \bibinfo {author} {\bibfnamefont {E.}~\bibnamefont {Bocquillon}}, \bibinfo
  {author} {\bibfnamefont {F.}~\bibnamefont {Dom\'{\i}nguez}}, \bibinfo
  {author} {\bibfnamefont {T.~M.}\ \bibnamefont {Klapwijk}}, \bibinfo {author}
  {\bibfnamefont {P.}~\bibnamefont {Leubner}}, \bibinfo {author} {\bibfnamefont
  {C.}~\bibnamefont {Br\"une}}, \bibinfo {author} {\bibfnamefont {E.~M.}\
  \bibnamefont {Hankiewicz}}, \bibinfo {author} {\bibfnamefont
  {S.}~\bibnamefont {Tarucha}}, \bibinfo {author} {\bibfnamefont
  {K.}~\bibnamefont {Ishibashi}}, \bibinfo {author} {\bibfnamefont
  {H.}~\bibnamefont {Buhmann}}, \ and\ \bibinfo {author} {\bibfnamefont
  {L.~W.}\ \bibnamefont {Molenkamp}},\ }\href {\doibase
  10.1103/PhysRevX.7.021011} {\bibfield  {journal} {\bibinfo  {journal} {Phys.
  Rev. X}\ }\textbf {\bibinfo {volume} {7}},\ \bibinfo {pages} {021011}
  (\bibinfo {year} {2017})}\BibitemShut {NoStop}%
\bibitem [{\citenamefont {Wiedenmann}\ \emph {et~al.}(2016)\citenamefont
  {Wiedenmann}, \citenamefont {Bocquillon}, \citenamefont {Deacon},
  \citenamefont {Hartinger}, \citenamefont {Herrmann}, \citenamefont
  {Klapwijk}, \citenamefont {Maier}, \citenamefont {Ames}, \citenamefont
  {Br\"une}, \citenamefont {Gould}, \citenamefont {Oiwa}, \citenamefont
  {Ishibashi}, \citenamefont {Tarucha}, \citenamefont {Buhmann},\ and\
  \citenamefont {Molenkamp}}]{wiedenmann:2016}%
  \BibitemOpen
  \bibfield  {author} {\bibinfo {author} {\bibfnamefont {J.}~\bibnamefont
  {Wiedenmann}}, \bibinfo {author} {\bibfnamefont {E.}~\bibnamefont
  {Bocquillon}}, \bibinfo {author} {\bibfnamefont {R.~S.}\ \bibnamefont
  {Deacon}}, \bibinfo {author} {\bibfnamefont {S.}~\bibnamefont {Hartinger}},
  \bibinfo {author} {\bibfnamefont {O.}~\bibnamefont {Herrmann}}, \bibinfo
  {author} {\bibfnamefont {T.~M.}\ \bibnamefont {Klapwijk}}, \bibinfo {author}
  {\bibfnamefont {L.}~\bibnamefont {Maier}}, \bibinfo {author} {\bibfnamefont
  {C.}~\bibnamefont {Ames}}, \bibinfo {author} {\bibfnamefont {C.}~\bibnamefont
  {Br\"une}}, \bibinfo {author} {\bibfnamefont {C.}~\bibnamefont {Gould}},
  \bibinfo {author} {\bibfnamefont {A.}~\bibnamefont {Oiwa}}, \bibinfo {author}
  {\bibfnamefont {K.}~\bibnamefont {Ishibashi}}, \bibinfo {author}
  {\bibfnamefont {S.}~\bibnamefont {Tarucha}}, \bibinfo {author} {\bibfnamefont
  {H.}~\bibnamefont {Buhmann}}, \ and\ \bibinfo {author} {\bibfnamefont
  {L.~W.}\ \bibnamefont {Molenkamp}},\ }\href
  {http://dx.doi.org/10.1038/ncomms10303} {\bibfield  {journal} {\bibinfo
  {journal} {Nature Communications}\ }\textbf {\bibinfo {volume} {7}},\
  \bibinfo {pages} {10303} (\bibinfo {year} {2016})}\BibitemShut {NoStop}%
\bibitem [{\citenamefont {Bocquillon}\ \emph {et~al.}(2016)\citenamefont
  {Bocquillon}, \citenamefont {Deacon}, \citenamefont {Wiedenmann},
  \citenamefont {Leubner}, \citenamefont {Klapwijk}, \citenamefont {Br\"une},
  \citenamefont {Ishibashi}, \citenamefont {Buhmann},\ and\ \citenamefont
  {Molenkamp}}]{bocquillon:2016}%
  \BibitemOpen
  \bibfield  {author} {\bibinfo {author} {\bibfnamefont {E.}~\bibnamefont
  {Bocquillon}}, \bibinfo {author} {\bibfnamefont {R.~S.}\ \bibnamefont
  {Deacon}}, \bibinfo {author} {\bibfnamefont {J.}~\bibnamefont {Wiedenmann}},
  \bibinfo {author} {\bibfnamefont {P.}~\bibnamefont {Leubner}}, \bibinfo
  {author} {\bibfnamefont {T.~M.}\ \bibnamefont {Klapwijk}}, \bibinfo {author}
  {\bibfnamefont {C.}~\bibnamefont {Br\"une}}, \bibinfo {author} {\bibfnamefont
  {K.}~\bibnamefont {Ishibashi}}, \bibinfo {author} {\bibfnamefont
  {H.}~\bibnamefont {Buhmann}}, \ and\ \bibinfo {author} {\bibfnamefont
  {L.~W.}\ \bibnamefont {Molenkamp}},\ }\href
  {http://dx.doi.org/10.1038/nnano.2016.159} {\bibfield  {journal} {\bibinfo
  {journal} {Nature Nanotechnology}\ }\textbf {\bibinfo {volume} {12}},\
  \bibinfo {pages} {137} (\bibinfo {year} {2016})}\BibitemShut {NoStop}%
\bibitem [{\citenamefont {van Woerkom}\ \emph {et~al.}(2017)\citenamefont {van
  Woerkom}, \citenamefont {Proutski}, \citenamefont {van Gulik}, \citenamefont
  {Kriv\'achy}, \citenamefont {Car}, \citenamefont {Plissard}, \citenamefont
  {Bakkers}, \citenamefont {Kouwenhoven},\ and\ \citenamefont
  {Geresdi}}]{woerkom:2017}%
  \BibitemOpen
  \bibfield  {author} {\bibinfo {author} {\bibfnamefont {D.~J.}\ \bibnamefont
  {van Woerkom}}, \bibinfo {author} {\bibfnamefont {A.}~\bibnamefont
  {Proutski}}, \bibinfo {author} {\bibfnamefont {R.~J.~J.}\ \bibnamefont {van
  Gulik}}, \bibinfo {author} {\bibfnamefont {T.}~\bibnamefont {Kriv\'achy}},
  \bibinfo {author} {\bibfnamefont {D.}~\bibnamefont {Car}}, \bibinfo {author}
  {\bibfnamefont {S.~R.}\ \bibnamefont {Plissard}}, \bibinfo {author}
  {\bibfnamefont {E.~P. A.~M.}\ \bibnamefont {Bakkers}}, \bibinfo {author}
  {\bibfnamefont {L.~P.}\ \bibnamefont {Kouwenhoven}}, \ and\ \bibinfo {author}
  {\bibfnamefont {A.}~\bibnamefont {Geresdi}},\ }\href {\doibase
  10.1103/PhysRevB.96.094508} {\bibfield  {journal} {\bibinfo  {journal} {Phys.
  Rev. B}\ }\textbf {\bibinfo {volume} {96}},\ \bibinfo {pages} {094508}
  (\bibinfo {year} {2017})}\BibitemShut {NoStop}%
\bibitem [{\citenamefont {Wang}\ \emph {et~al.}(2016)\citenamefont {Wang},
  \citenamefont {Deacon}, \citenamefont {Car}, \citenamefont {Bakkers},\ and\
  \citenamefont {Ishibashi}}]{wang:2016}%
  \BibitemOpen
  \bibfield  {author} {\bibinfo {author} {\bibfnamefont {R.}~\bibnamefont
  {Wang}}, \bibinfo {author} {\bibfnamefont {R.~S.}\ \bibnamefont {Deacon}},
  \bibinfo {author} {\bibfnamefont {D.}~\bibnamefont {Car}}, \bibinfo {author}
  {\bibfnamefont {E.~P. A.~M.}\ \bibnamefont {Bakkers}}, \ and\ \bibinfo
  {author} {\bibfnamefont {K.}~\bibnamefont {Ishibashi}},\ }\href {\doibase
  10.1063/1.4950764} {\bibfield  {journal} {\bibinfo  {journal} {Applied
  Physics Letters}\ }\textbf {\bibinfo {volume} {108}},\ \bibinfo {pages}
  {203502} (\bibinfo {year} {2016})}\BibitemShut {NoStop}%
\bibitem [{\citenamefont {Baba}\ \emph {et~al.}(2017)\citenamefont {Baba},
  \citenamefont {Matsuo}, \citenamefont {Kamata}, \citenamefont {Deacon},
  \citenamefont {Oiwa}, \citenamefont {Li}, \citenamefont {Jeppesen},
  \citenamefont {Samuelson}, \citenamefont {Xu},\ and\ \citenamefont
  {Tarucha}}]{baba:2017}%
  \BibitemOpen
  \bibfield  {author} {\bibinfo {author} {\bibfnamefont {S.}~\bibnamefont
  {Baba}}, \bibinfo {author} {\bibfnamefont {S.}~\bibnamefont {Matsuo}},
  \bibinfo {author} {\bibfnamefont {H.}~\bibnamefont {Kamata}}, \bibinfo
  {author} {\bibfnamefont {R.~S.}\ \bibnamefont {Deacon}}, \bibinfo {author}
  {\bibfnamefont {A.}~\bibnamefont {Oiwa}}, \bibinfo {author} {\bibfnamefont
  {K.}~\bibnamefont {Li}}, \bibinfo {author} {\bibfnamefont {S.}~\bibnamefont
  {Jeppesen}}, \bibinfo {author} {\bibfnamefont {L.}~\bibnamefont {Samuelson}},
  \bibinfo {author} {\bibfnamefont {H.~Q.}\ \bibnamefont {Xu}}, \ and\ \bibinfo
  {author} {\bibfnamefont {S.}~\bibnamefont {Tarucha}},\ }\href {\doibase
  10.1063/1.4997646} {\bibfield  {journal} {\bibinfo  {journal} {Applied
  Physics Letters}\ }\textbf {\bibinfo {volume} {111}},\ \bibinfo {pages}
  {233513} (\bibinfo {year} {2017})}\BibitemShut {NoStop}%
\bibitem [{\citenamefont {Moshchalkov}\ \emph {et~al.}(1995)\citenamefont
  {Moshchalkov}, \citenamefont {Gielen}, \citenamefont {Strunk}, \citenamefont
  {Jonckheere}, \citenamefont {Qiu}, \citenamefont {Haesendonck},\ and\
  \citenamefont {Bruynseraede}}]{moshchalkov:1995}%
  \BibitemOpen
  \bibfield  {author} {\bibinfo {author} {\bibfnamefont {V.~V.}\ \bibnamefont
  {Moshchalkov}}, \bibinfo {author} {\bibfnamefont {L.}~\bibnamefont {Gielen}},
  \bibinfo {author} {\bibfnamefont {C.}~\bibnamefont {Strunk}}, \bibinfo
  {author} {\bibfnamefont {R.}~\bibnamefont {Jonckheere}}, \bibinfo {author}
  {\bibfnamefont {X.}~\bibnamefont {Qiu}}, \bibinfo {author} {\bibfnamefont
  {C.~V.}\ \bibnamefont {Haesendonck}}, \ and\ \bibinfo {author} {\bibfnamefont
  {Y.}~\bibnamefont {Bruynseraede}},\ }\href
  {http://dx.doi.org/10.1038/373319a0} {\bibfield  {journal} {\bibinfo
  {journal} {Nature}\ }\textbf {\bibinfo {volume} {373}},\ \bibinfo {pages}
  {319} (\bibinfo {year} {1995})}\BibitemShut {NoStop}%
\bibitem [{\citenamefont {Poza}\ \emph {et~al.}(1998)\citenamefont {Poza},
  \citenamefont {Bascones}, \citenamefont {Rodrigo}, \citenamefont
  {Agra\"{\i}t}, \citenamefont {Vieira},\ and\ \citenamefont
  {Guinea}}]{poza:1998}%
  \BibitemOpen
  \bibfield  {author} {\bibinfo {author} {\bibfnamefont {M.}~\bibnamefont
  {Poza}}, \bibinfo {author} {\bibfnamefont {E.}~\bibnamefont {Bascones}},
  \bibinfo {author} {\bibfnamefont {J.~G.}\ \bibnamefont {Rodrigo}}, \bibinfo
  {author} {\bibfnamefont {N.}~\bibnamefont {Agra\"{\i}t}}, \bibinfo {author}
  {\bibfnamefont {S.}~\bibnamefont {Vieira}}, \ and\ \bibinfo {author}
  {\bibfnamefont {F.}~\bibnamefont {Guinea}},\ }\href {\doibase
  10.1103/PhysRevB.58.11173} {\bibfield  {journal} {\bibinfo  {journal} {Phys.
  Rev. B}\ }\textbf {\bibinfo {volume} {58}},\ \bibinfo {pages} {11173}
  (\bibinfo {year} {1998})}\BibitemShut {NoStop}%
\bibitem [{\citenamefont {Suyatin}\ \emph {et~al.}(2007)\citenamefont
  {Suyatin}, \citenamefont {Thelander}, \citenamefont {Bj\"{o}rk},
  \citenamefont {Maximov},\ and\ \citenamefont {Samuelson}}]{suyatin:2007}%
  \BibitemOpen
  \bibfield  {author} {\bibinfo {author} {\bibfnamefont {D.~B.}\ \bibnamefont
  {Suyatin}}, \bibinfo {author} {\bibfnamefont {C.}~\bibnamefont {Thelander}},
  \bibinfo {author} {\bibfnamefont {M.~T.}\ \bibnamefont {Bj\"{o}rk}}, \bibinfo
  {author} {\bibfnamefont {I.}~\bibnamefont {Maximov}}, \ and\ \bibinfo
  {author} {\bibfnamefont {L.}~\bibnamefont {Samuelson}},\ }\href
  {http://stacks.iop.org/0957-4484/18/i=10/a=105307} {\bibfield  {journal}
  {\bibinfo  {journal} {Nanotechnology}\ }\textbf {\bibinfo {volume} {18}},\
  \bibinfo {pages} {105307} (\bibinfo {year} {2007})}\BibitemShut {NoStop}%
\bibitem [{\citenamefont {G\"ul}\ \emph {et~al.}(2017)\citenamefont {G\"ul},
  \citenamefont {Zhang}, \citenamefont {de~Vries}, \citenamefont {van Veen},
  \citenamefont {Zuo}, \citenamefont {Mourik}, \citenamefont {Conesa-Boj},
  \citenamefont {Nowak}, \citenamefont {van Woerkom}, \citenamefont
  {Quintero-P\'erez}, \citenamefont {Cassidy}, \citenamefont {Geresdi},
  \citenamefont {Koelling}, \citenamefont {Car}, \citenamefont {Plissard},
  \citenamefont {Bakkers},\ and\ \citenamefont {Kouwenhoven}}]{gul:2017}%
  \BibitemOpen
  \bibfield  {author} {\bibinfo {author} {\bibfnamefont {O.}~\bibnamefont
  {G\"ul}}, \bibinfo {author} {\bibfnamefont {H.}~\bibnamefont {Zhang}},
  \bibinfo {author} {\bibfnamefont {F.~K.}\ \bibnamefont {de~Vries}}, \bibinfo
  {author} {\bibfnamefont {J.}~\bibnamefont {van Veen}}, \bibinfo {author}
  {\bibfnamefont {K.}~\bibnamefont {Zuo}}, \bibinfo {author} {\bibfnamefont
  {V.}~\bibnamefont {Mourik}}, \bibinfo {author} {\bibfnamefont
  {S.}~\bibnamefont {Conesa-Boj}}, \bibinfo {author} {\bibfnamefont {M.~P.}\
  \bibnamefont {Nowak}}, \bibinfo {author} {\bibfnamefont {D.~J.}\ \bibnamefont
  {van Woerkom}}, \bibinfo {author} {\bibfnamefont {M.}~\bibnamefont
  {Quintero-P\'erez}}, \bibinfo {author} {\bibfnamefont {M.~C.}\ \bibnamefont
  {Cassidy}}, \bibinfo {author} {\bibfnamefont {A.}~\bibnamefont {Geresdi}},
  \bibinfo {author} {\bibfnamefont {S.}~\bibnamefont {Koelling}}, \bibinfo
  {author} {\bibfnamefont {D.}~\bibnamefont {Car}}, \bibinfo {author}
  {\bibfnamefont {S.~R.}\ \bibnamefont {Plissard}}, \bibinfo {author}
  {\bibfnamefont {E.~P. A.~M.}\ \bibnamefont {Bakkers}}, \ and\ \bibinfo
  {author} {\bibfnamefont {L.~P.}\ \bibnamefont {Kouwenhoven}},\ }\href
  {\doibase 10.1021/acs.nanolett.7b00540} {\bibfield  {journal} {\bibinfo
  {journal} {Nano Letters}\ }\textbf {\bibinfo {volume} {17}},\ \bibinfo
  {pages} {2690} (\bibinfo {year} {2017})}\BibitemShut {NoStop}%
\bibitem [{\citenamefont {G\"ul}\ \emph {et~al.}(2018)\citenamefont {G\"ul},
  \citenamefont {Zhang}, \citenamefont {Bommer}, \citenamefont {de~Moor},
  \citenamefont {Car}, \citenamefont {Plissard}, \citenamefont {Bakkers},
  \citenamefont {Geresdi}, \citenamefont {Watanabe}, \citenamefont
  {Taniguchi},\ and\ \citenamefont {Kouwenhoven}}]{gul:2018}%
  \BibitemOpen
  \bibfield  {author} {\bibinfo {author} {\bibfnamefont {O.}~\bibnamefont
  {G\"ul}}, \bibinfo {author} {\bibfnamefont {H.}~\bibnamefont {Zhang}},
  \bibinfo {author} {\bibfnamefont {J.~D.~S.}\ \bibnamefont {Bommer}}, \bibinfo
  {author} {\bibfnamefont {M.~W.~A.}\ \bibnamefont {de~Moor}}, \bibinfo
  {author} {\bibfnamefont {D.}~\bibnamefont {Car}}, \bibinfo {author}
  {\bibfnamefont {S.~R.}\ \bibnamefont {Plissard}}, \bibinfo {author}
  {\bibfnamefont {E.~P. A.~M.}\ \bibnamefont {Bakkers}}, \bibinfo {author}
  {\bibfnamefont {A.}~\bibnamefont {Geresdi}}, \bibinfo {author} {\bibfnamefont
  {K.}~\bibnamefont {Watanabe}}, \bibinfo {author} {\bibfnamefont
  {T.}~\bibnamefont {Taniguchi}}, \ and\ \bibinfo {author} {\bibfnamefont
  {L.~P.}\ \bibnamefont {Kouwenhoven}},\ }\href {\doibase
  10.1038/s41565-017-0032-8} {\bibfield  {journal} {\bibinfo  {journal} {Nature
  Nanotechnology}\ } (\bibinfo {year} {2018}),\
  10.1038/s41565-017-0032-8}\BibitemShut {NoStop}%
\bibitem [{\citenamefont {Chauvin}\ \emph {et~al.}(2006)\citenamefont
  {Chauvin}, \citenamefont {vom Stein}, \citenamefont {Pothier}, \citenamefont
  {Joyez}, \citenamefont {Huber}, \citenamefont {Esteve},\ and\ \citenamefont
  {Urbina}}]{chauvin:2006}%
  \BibitemOpen
  \bibfield  {author} {\bibinfo {author} {\bibfnamefont {M.}~\bibnamefont
  {Chauvin}}, \bibinfo {author} {\bibfnamefont {P.}~\bibnamefont {vom Stein}},
  \bibinfo {author} {\bibfnamefont {H.}~\bibnamefont {Pothier}}, \bibinfo
  {author} {\bibfnamefont {P.}~\bibnamefont {Joyez}}, \bibinfo {author}
  {\bibfnamefont {M.~E.}\ \bibnamefont {Huber}}, \bibinfo {author}
  {\bibfnamefont {D.}~\bibnamefont {Esteve}}, \ and\ \bibinfo {author}
  {\bibfnamefont {C.}~\bibnamefont {Urbina}},\ }\href {\doibase
  10.1103/PhysRevLett.97.067006} {\bibfield  {journal} {\bibinfo  {journal}
  {Phys. Rev. Lett.}\ }\textbf {\bibinfo {volume} {97}},\ \bibinfo {pages}
  {067006} (\bibinfo {year} {2006})}\BibitemShut {NoStop}%
\bibitem [{\citenamefont {Shapiro}(1963)}]{shapiro:1963}%
  \BibitemOpen
  \bibfield  {author} {\bibinfo {author} {\bibfnamefont {S.}~\bibnamefont
  {Shapiro}},\ }\href {\doibase 10.1103/PhysRevLett.11.80} {\bibfield
  {journal} {\bibinfo  {journal} {Phys. Rev. Lett.}\ }\textbf {\bibinfo
  {volume} {11}},\ \bibinfo {pages} {80} (\bibinfo {year} {1963})}\BibitemShut
  {NoStop}%
\bibitem [{\citenamefont {Reale}(1974)}]{reale:1974}%
  \BibitemOpen
  \bibfield  {author} {\bibinfo {author} {\bibfnamefont {C.}~\bibnamefont
  {Reale}},\ }\href {\doibase 10.1007/BF03157926} {\bibfield  {journal}
  {\bibinfo  {journal} {Acta Physica Academiae Scientiarum Hungaricae}\
  }\textbf {\bibinfo {volume} {37}},\ \bibinfo {pages} {53} (\bibinfo {year}
  {1974})}\BibitemShut {NoStop}%
\end{thebibliography}

\begin{thebibliography}{2}%
\makeatletter
\providecommand \@ifxundefined [1]{%
 \@ifx{#1\undefined}
}%
\providecommand \@ifnum [1]{%
 \ifnum #1\expandafter \@firstoftwo
 \else \expandafter \@secondoftwo
 \fi
}%
\providecommand \@ifx [1]{%
 \ifx #1\expandafter \@firstoftwo
 \else \expandafter \@secondoftwo
 \fi
}%
\providecommand \natexlab [1]{#1}%
\providecommand \enquote  [1]{``#1''}%
\providecommand \bibnamefont  [1]{#1}%
\providecommand \bibfnamefont [1]{#1}%
\providecommand \citenamefont [1]{#1}%
\providecommand \href@noop [0]{\@secondoftwo}%
\providecommand \href [0]{\begingroup \@sanitize@url \@href}%
\providecommand \@href[1]{\@@startlink{#1}\@@href}%
\providecommand \@@href[1]{\endgroup#1\@@endlink}%
\providecommand \@sanitize@url [0]{\catcode `\\12\catcode `\$12\catcode
  `\&12\catcode `\#12\catcode `\^12\catcode `\_12\catcode `\%12\relax}%
\providecommand \@@startlink[1]{}%
\providecommand \@@endlink[0]{}%
\providecommand \url  [0]{\begingroup\@sanitize@url \@url }%
\providecommand \@url [1]{\endgroup\@href {#1}{\urlprefix }}%
\providecommand \urlprefix  [0]{URL }%
\providecommand \Eprint [0]{\href }%
\providecommand \doibase [0]{http://dx.doi.org/}%
\providecommand \selectlanguage [0]{\@gobble}%
\providecommand \bibinfo  [0]{\@secondoftwo}%
\providecommand \bibfield  [0]{\@secondoftwo}%
\providecommand \translation [1]{[#1]}%
\providecommand \BibitemOpen [0]{}%
\providecommand \bibitemStop [0]{}%
\providecommand \bibitemNoStop [0]{.\EOS\space}%
\providecommand \EOS [0]{\spacefactor3000\relax}%
\providecommand \BibitemShut  [1]{\csname bibitem#1\endcsname}%
\let\auto@bib@innerbib\@empty
\bibitem [{\citenamefont {Olsson}\ \emph {et~al.}(1996)\citenamefont {Olsson},
  \citenamefont {Andersson}, \citenamefont {H\aa{}kansson}, \citenamefont
  {Kanski}, \citenamefont {Ilver},\ and\ \citenamefont
  {Karlsson}}]{olsson:1996}%
  \BibitemOpen
  \bibfield  {author} {\bibinfo {author} {\bibfnamefont {L.~O.}\ \bibnamefont
  {Olsson}}, \bibinfo {author} {\bibfnamefont {C.~B.~M.}\ \bibnamefont
  {Andersson}}, \bibinfo {author} {\bibfnamefont {M.~C.}\ \bibnamefont
  {H\aa{}kansson}}, \bibinfo {author} {\bibfnamefont {J.}~\bibnamefont
  {Kanski}}, \bibinfo {author} {\bibfnamefont {L.}~\bibnamefont {Ilver}}, \
  and\ \bibinfo {author} {\bibfnamefont {U.~O.}\ \bibnamefont {Karlsson}},\
  }\href {\doibase 10.1103/PhysRevLett.76.3626} {\bibfield  {journal} {\bibinfo
   {journal} {Phys. Rev. Lett.}\ }\textbf {\bibinfo {volume} {76}},\ \bibinfo
  {pages} {3626} (\bibinfo {year} {1996})}\BibitemShut {NoStop}%
\bibitem [{\citenamefont {Dayem}\ and\ \citenamefont
  {Grimes}(1966)}]{dayem:1966}%
  \BibitemOpen
  \bibfield  {author} {\bibinfo {author} {\bibfnamefont {A.~H.}\ \bibnamefont
  {Dayem}}\ and\ \bibinfo {author} {\bibfnamefont {C.~C.}\ \bibnamefont
  {Grimes}},\ }\href {\doibase 10.1063/1.1754595} {\bibfield  {journal}
  {\bibinfo  {journal} {Applied Physics Letters}\ }\textbf {\bibinfo {volume}
  {9}},\ \bibinfo {pages} {47} (\bibinfo {year} {1966})}\BibitemShut {NoStop}%
\end{thebibliography}
\end{document}